\documentclass[universe,article,accept,moreauthors,pdftex]{Definitions/mdpi} 
\usepackage{bm,amsfonts}
\usepackage{amssymb}
\usepackage{cancel}
 \usepackage{subfig}
 \usepackage{siunitx}
 \usepackage{microtype}
\usepackage{textcomp}
\firstpage{1} 
\pubvolume{7}
\issuenum{1}
\articlenumber{234}
\pubyear{2021}
\copyrightyear{2020}
\datereceived{30 May 2001} 
\dateaccepted{5 July 2001} 
\datepublished{9 July 2001} 
\hreflink{https://doi.org/} 
\pdfoutput=1
\newcommand{\eqnref}[1]{Eqn.~\mbox{(\ref{#1})}}		
\newcommand{\eqnsref}[1]{Eqns.~\mbox{(\ref{#1})}}		
\newcommand{\figref}[1]{Fig.~\ref{#1}}			
\newcommand{\tabref}[1]{Tab.~\ref{#1}}			
\newcommand{\secref}[1]{Section~\ref{#1}}		
\newcommand{\appref}[1]{Appendix~\ref{#1}}		
\newcommand{\pa}{\partial}						
\newcommand{\diff}[2]{\frac{\pa #1}{\pa #2}}				
\newcommand{\cannex}{{\sc{Cannex}}}				
\newcommand{\ri}{{\rm i}}						
\newcommand{\re}{{\rm e}}						
\renewcommand{\a}{\alpha}

\newcommand{\vare}{\varepsilon}

\newcommand{\x}{\xi}
\renewcommand{\r}{\rho}

\newcommand{\inv}[1]{\frac{1}{#1}}					
\graphicspath{{./}{./figures/}{./Definitions/}}
\Title{Next generation design and prospects for \cannex{}}

\TitleCitation{Next generation design and prospects for \cannex{}}

\Author{Ren\'e I.P. Sedmik$^{1,\dagger}$\orcidA{} and Mario Pitschmann $^{2,\dagger}$\orcidB{}}

\AuthorNames{Rene I.P. Sedmik and Mario Pitschmann}

\AuthorCitation{Sedmik, R. I.P.; Pitschmann, M.}

\address{%
$^{1}$ \quad Correspondence: rene.sedmik@tuwien.ac.at\\
$^{2}$ \quad mario.pitschmann@tuwien.ac.at}

\firstnote{Current address: Technische Universit\"{a}t Wien, Atominstitut, Stadionallee 2, 1020 Vienna, Austria} 
 \abstract{The Casimir And Non-Newtonian force EXperiment (\cannex{}) implements the unique geometry of macroscopic plane parallel plates that guarantees an optimum sensitivity with respect to interfacial forces and their gradients. Based on experience from the recently completed proof-of-principle phase, we have started a re-design of the setup aiming to reduce systematic effects and maximize the achievable sensitivity. Several propositions have been made to measure Casimir forces in and out of thermal equilibrium, hypothetical axion and axion-like dark matter interactions, and forces originating from chameleon or symmetron dark energy interactions. In the present article, we give details on the design for the next implementation stage of \cannex{} and discuss the experimental opportunities as well as limitations expected for this new setup.}

\keyword{quantum vacuum; Casimir pressure; axion; non-Newtonian gravity}

\begin{document}
%
%
%
\section{Introduction}
\label{sec:intro}
A century after the inception of general relativity (GR) and quantum mechanics (QM), these two pillars of our modern understanding of physics have evolved into the standard model of particle physics (SM) and the cosmological standard model $\Lambda$-CDM. In recent years, however, experiments have produced results that are difficult to interpret within this framework, which suggests that the latter may be incomplete~\cite{Bull:2015stt}. Indeed, a series of newly discovered discrepancies concerning the Hubble constant and $\sigma_8$ parameter are difficult to reconcile with the current theoretical framework. The Hubble constant obtained from large redshift Planck data~\cite{Ade:2015xua,Aghanim:2018eyx} is at tension~\cite{Raveri:2018wln} with the one extracted from baryoacoustic oscillation data at the level of $4.4\sigma$~\cite{Riess:2019cxk}, or even more~\cite{Jee:2019hah,DiValentino:2019qzk}. The $\sigma_8$ parameter is afflicted by a similar tension~\cite{Battye:2014qga}. Generically, it seems that large redshift measurements (Planck) do not agree with those measured in the local Universe (weak lensing and clustering~\cite{Abbott:2017wau}, and Cepheid~\cite{Riess:2019cxk} data). Still another conflict arises from the Planck 2018 data~\cite{Aghanim:2018eyx}, which seem to favor finite curvature~\cite{DiValentino:2019qzk} contrary to what is expected from cosmic inflation. Eventually, the observation of galaxies with too little dark mass also excludes $\Lambda$-CDM by $\geq 4.8\sigma$~\cite{Haslbauer:2019cpl}. By now, systematic errors have convincingly been ruled out as the source of these discrepancies. It, thus, appears that $\Lambda$-CDM, which only considers a global Hubble constant and a flat Universe, requires some modification. With respect to the SM, indications for the existence of sterile neutrinos are given by significant deviations from expectations of the measured neutrino fluxes in dedicated experiments~\cite{Aguilar-Arevalo:2018gpe,Aguilar:2001ty}, at reactors~\cite{Mention:2011rk} (for some remaining inconsistencies, see Reference~\cite{Dentler:2018sju}), and around chemical sources~\mbox{\cite{Barinov:2017ymq,Kaether:2010ag,Abdurashitov:2005tb}}. Unexpected emission lines in astronomical X-ray data may be also be interpreted as signs of an unknown neutrino species~\cite{Bulbul:2014sua,Hofmann:2019ihc}. This supports the view that also the SM is in fact incomplete and that new light particles yet to be found might exist.

These discrepancies add to existing long-standing problems, such as the cosmological constant problem~\cite{Weinberg:1988cp}, strong CP-violation problem~\cite{Dine:2000cj}, or the hierarchy problem~\cite{Burdman:2007,Graham:2015cka}, which are clear indications for our understanding of cosmic evolution and the quantum vacuum to be incomplete. Observations, in this respect, however, are not limited to astronomical measurements, but insight can also be gained in the laboratory. In high-energy physics, with the experimental detection of the Higgs particle, a compelling mechanism for the generation of mass was verified. Most of the candidate fields and interactions proposed to represent the dark sector, however, have much lower (sub-eV) masses and may be inaccessible to direct measurements in colliders. An alternative way to detect the minuscule effects of proposed dark sector models is to search for deviations from Newton's law at small separations. Benefiting from technological progress, thus, metrological force experiments have gained importance. Traditionally, in the range \SI{10}{\nano\metre} to \SI{1}{\milli\metre} of separations, Cavendish-type experiments~\cite{Adelberger:2003zx,Hoyle:2004cw,Kapner:2006si,Geraci:2008hb,Adelberger:2009zz,Hoedl:2011zz,Heckel:2013ina,Terrano:2015sna,Lee:2020zjt,Tan:2020vpf,Zhao:2021anp} and Casimir experiments~\cite{Masuda:2009vu,Sushkov:2011md,Decca:2003td,Decca:2005yk,Decca:2007jq,Bezerra:2010pq,Klimchitskaya:2012pc,Chen:2014oda,Bimonte:2021} are most sensitive (see Refs.~\cite{Sponar:2020bhj,Safronova:2017xyt,Klimchitskaya:2021gkd} for more complete reviews). The Casimir And Non-Newtonian force EXperiment (\cannex{}) was designed from the start to measure interfacial \emph{and} gravity-like interactions~\cite{Brax:2010xx,Almasi:2015zpa} between macroscopic parallel plates. Recently, the experimental proof of principle was accomplished~\cite{Sedmik:2018kqt}, which motivated a number of further propositions for possible measurements, ranging from Casimir forces~\cite{Klimchitskaya:2019fsm} in and out of thermal equilibrium~\cite{Klimchitskaya:2019nzu}, to dark energy chameleon forces, as well as scalar and vector-valued Yukawa interactions~\cite{Klimchitskaya:2019fsm,Pitschmann:2021}. In the present paper, we first review the theoretical background, experimental evidence, and open questions regarding dark sector interactions and Casimir physics in Sections \ref{sec:dark} and \ref{sec:vacuum}, respectively. A generic motivation for the use of the parallel plate geometry in force metrology is given in \secref{sec:pp}. These first sections may serve as a self-consistent comprehensive review surveying all topics of relevance for \cannex{}. Readers interested in the experiment may skip directly to \secref{sec:design}, 
where we discuss technical improvements to be introduced in the upcoming next version of \cannex{} and give a detailed account of the realistically expected errors and the resulting sensitivity. Based on these error calculations, we update several prospective results in \secref{sec:prospects} and conclude our findings in \secref{sec:conclusion}.

\section{Overview Dark Sector Interactions}
\label{sec:dark}
In the following, we provide a brief introduction to the dark sector (for other recent reviews along similar lines, also see Reference~\cite{Sponar:2020bhj,Klimchitskaya:2021gkd}). As mentioned previously, our knowledge of fundamental physics is based on the SM and GR in terms of the $\Lambda$-CDM model. While the former, with the recent detection of the Higgs particle, has been enormously successful in describing particle phenomenology, it leaves many questions unanswered. Similarly, GR has been incredibly successful in accounting for ever more experimental tests and theoretical investigations~\cite{Will:2014kxa}. Despite those successes, there are strong reasons to believe that GR is incomplete at high energies and will be supplanted by a more fundamental theory. Despite intensive efforts for almost a century by many of the most distinguished physicists, GR has resisted any attempt at its quantization, which is in strong contrast to all other known fundamental interactions (notwithstanding the progress made by string theory, loop quantum gravity and other approaches to quantum gravity). As we now know, the particle content of the SM provides only a very minor fraction of the matter/energy content of our present Universe. By far, the larger part lies in the `dark', consisting of dark matter (DM) and dark energy (DE), as has been revealed by the Planck Mission study of the cosmic microwave background radiation~\cite{Adam:2015rua}. Specifically, the current matter content of the Universe consists of 69\% dark energy, 26\% dark matter, and only 5\% SM matter. 

\subsection{Dark Matter}
\label{sec:ODSIDM}
The (pre-)history of dark matter reaches back into the early 20th century and may be traced further to investigations done by Lord Kelvin~\cite{Thomson:2010} and Poincar\'e~\cite{hp1906sa} (see Reference~\cite{Bertone:2016nfn} for an account of the history of dark matter). In the 1930s, Zwicky postulated the existence of dark matter after studying galaxy clusters and obtaining more concrete indications of unseen mass~\cite{Zwicky:1933, Zwicky:1937zza}. After a long history of growing evidence, the existence of dark matter is now well established through numerous studies of astronomical objects~\cite{Bertone:2010zza}. Nevertheless, its origin is still unknown, with the only certainty being that it is not to be found within the SM. On the other hand, the SM leaves many vexing questions unanswered, which continues to inspire a plethora of proposed generalizations and extensions. The latter typically postulate the existence of hypothetical new particles, which may comprise possible candidates for dark matter particles (as, e.g., the non-SM sterile neutrinos mentioned above). A particularly well motivated example has been devised by Peccei and Quinn~\cite{Peccei:1977hh, Peccei:1977ur} in order to solve the so-called strong $CP$ problem. The corresponding hypothetical new pseudo-scalar particle has been termed `axion'~\cite{Weinberg:1977ma, Wilczek:1977pj, Kim:1979if, Shifman:1979if, Dine:1981rt}. For comprehensive and detailed reviews on axions, we refer to Reference~\cite{Pospelov:2005pr,Kim:2008hd}; for a compact survey of its mechanism and a comparison to ALPs, see Reference~\cite{Mantry:2014zsa}; and a more up-to-date compact survey can be found in Reference~\cite{10.1093/ptep/ptaa104}. The corresponding theory has been developed and extended vastly since their original formulation by Peccei and Quinn. In the following, we confine the discussion to the original QCD axion, which mediates a force via a light neutral spin-0 particle $\phi$ that interacts with quarks of flavor $q=u,d$ through the interaction Lagrangian
 \begin{align}
\label{eq:axionquark}
{\cal L}_{\phi qq} = \phi\sum\limits_{q=u,d} \>\bar{q} \big(g_S^q + ig_P^q\gamma^5\big) q\>,
\end{align} 
with axion-quark scalar $g_S^q$ and pseudo-scalar $g_P^q$ coupling constants and the Dirac matrix $\gamma^5=i\gamma^0\gamma^1\gamma^2\gamma^3$. Thereby, the effective axion-nucleon ($N = p,n$) interaction Lagrangian is induced at the hadronic level
\begin{align}
\label{eq:axionnucleon}
{\cal L}_{\phi NN} = \phi\sum\limits_{N=p,n}  \>\bar{N}\big(g_S + ig_P\gamma^5 \big) N\>,
\end{align}
with the scalar $g_S$ and pseudo-scalar $g_P$ coupling constants (where we have assumed purely isoscalar interactions $g_{S,P}^u=g_{S,P}^d$ for simplicity). From the Lagrangian \eqnref{eq:axionnucleon}, the non-relativistic effective potentials describing the axion-induced interactions between nucleons are obtained:
\begin{align}
V_{SS}(r)&=-\frac{g_{S,1}g_{S,2}\hbar c}{4 \pi}\frac{\re^{-r/\lambda}}{r}\quad &&\text{scalar},\label{eq:scalar-scalar}\\
 V_{PS}(r)&=-\frac{g_{P,1} g_{S,2} \hbar}{4\pi m_1}\,\bm{s}_1\cdot\hat{\bf{r}}\,\left(\inv{r}+\inv{\lambda}\right)\,\frac{\re^{-r/\lambda}}{r} &&\text{pseudoscalar-scalar}.\label{eq:pseudoscalar-scalar}
 \end{align}
Here, $\bm{s}_1=(\hbar/2)\bm{\sigma}_1$, while $m_1$ and $\bm{\sigma}_1$ denote the mass and Pauli spin-matrix vector of nucleon $1$, respectively, $\hat{\bf{r}} = {\bf{r}}/r$ is the unit vector pointing from nucleon $2$ to nucleon $1$, and $\lambda = \hbar/(m_\phi c)$ is the interaction range for an axion mass $m_\phi$ (in the more general case of two different vertices for the potential \eqnref{eq:pseudoscalar-scalar}, one has to add the same potential with exchanged indices $(1\leftrightarrow 2)$ due to permutation symmetry). These effective potentials have been obtained for the first time in Reference~\cite{Moody:1984ba} (for more recent work pointing out several issues of earlier investigations, see Reference~\cite{Fadeev:2018rfl}).

Next, we will succinctly summarize how the Peccei-Quinn mechanism solves the strong $CP$ problem~\cite{Peccei:1977hh,Peccei:1977ur,Weinberg:1977ma,Wilczek:1977pj}. The SM has two possible sources of $CP$ violation leading to non-vanishing electric dipole moments (EDMs); one is due to the complex phase in the CKM matrix characterizing the strength of flavor-changing charged currents. However, the corresponding neutron EDM is many orders of magnitude below the current experimental limit, rendering it irrelevant, at present. The second $CP$ violating source is due to the term 
\begin{align}
\label{thetaterm}
{\cal L}_{QCD}^{\cancel{CP}} = \bar{\theta}\> \frac{\alpha_s}{16 \pi}\> G_{\mu \nu}^a \tilde{G}^{a,\mu \nu}\>,
\end{align}
in the QCD Lagrangian. Here, $\tilde{G}_{\mu \nu}^a=\varepsilon_{\mu\nu\rho\sigma} G^{a,\rho \sigma}$ is the dual chromodynamic field strength tensor, and the parameter $\bar{\theta}$ is given by
\begin{align}
\label{thetabar}
\bar{\theta} = \theta + \text{arg}(\text{det} \>M_q')\>,
\end{align}
and it arises from the non-trivial structure of the QCD vacuum. $M_q'$ is the original, still non-diagonal, quark mass matrix after electroweak symmetry breaking. Since the term in \eqnref{thetaterm} is not forbidden by any symmetry, it is actually expected due to the non-trivial structure of the QCD vacuum, the axial $U(1)_A$ anomaly~\cite{Adler:1969gk, Bell:1969ts}, and the absence of massless quarks in the SM. In contrast to the flavor-changing $CP$ violation due to the CKM phase, the $\bar{\theta}$-term leads to a flavor-diagonal $CP$ violation.
Experimental EDM limits on neutrons and mercury put stringent bounds on this parameter, i.e.,
\begin{align}
|\>\bar{\theta}\>| \lesssim 10^{-11}\>. 
\end{align}

Expectations on dimensional grounds suggest $\theta\sim\mathcal O(1)$. In addition, further contributions to $\theta$ arise from the electroweak sector, as well as from radiative corrections. All these contributions combine to an effective $\bar\theta$. Hence, the strong $CP$ problem amounts to the question as to why several contributions, coming from unrelated sources in the SM, all add up to a total effective contribution many orders of magnitude smaller than its estimated individual contributions. 

A natural solution to this problem is provided by the axion mechanism~\cite{Peccei:1977hh}. Among the host of different axion models, we discuss the Kim-Shifman-Vainstein-Zakharov (KSVZ) model~\cite{Kim:1979if,Shifman:1979if}, in which the SM is augmented by a new massless electroweak-singlet quark $\psi$ and a complex scalar $\Phi$. The corresponding Lagrangian, and with it the entire SM, is invariant under a global chiral $U(1)_{\text{PQ}}$ Peccei-Quinn transformation 
\begin{align}
\label{PQ}
\psi \to e^{-i\alpha \gamma^5}\psi\>, \qquad \bar\psi\to\bar\psi\,e^{-i\alpha \gamma^5}\>, \qquad  \Phi \to e^{-2i\alpha} \Phi\>.
\end{align}

The Peccei-Quinn transformation is anomalous and contributes a shift to the value of $\bar{\theta}$, analogously to an axial $U(1)_A$ transformation. Consequently, one can eliminate the $\bar{\theta}$-parameter by an appropriate $U(1)_{\text{PQ}}$ rotation, which solves the strong $CP$ problem. A massless quark has not been observed in nature. Therefore, the $U(1)_{\text{PQ}}$ symmetry of the Lagrangian must be spontaneously broken at a high enough Peccei-Quinn scale of \mbox{$f_a \sim 10^9$--$10^{12}\,$GeV} in order for the new quark to acquire a mass large enough to render it invisible within current experimental limits. Since the $U(1)_{\text{PQ}}$ symmetry is explicitly broken by the chiral anomaly, the axion, being the \emph{pseudo}-Goldstone boson associated with the corresponding spontaneous symmetry breaking, acquires a potential and a non-zero mass. An effective theory for low energies is obtained by integrating out the heavy fields $\psi$ with the SM fields and the axion being left as the surviving low energy degrees of freedom.
Effectively, in the Peccei-Quinn mechanism, $\bar\theta$ is promoted to a dynamical field, which acquires a small but non-vanishing minimum value $\theta_{\text{eff}}$ consistent with experimental bounds, due to its minimization in the corresponding ground state axion potential. Particularly beautiful is the explicit prediction of the axion mass and of its couplings in \eqnref{eq:axionquark} in terms of the breaking scale $f_a$, quark masses $m_q$, and effective minimum value $\theta_{\text{eff}}$~\cite{Mantry:2014zsa},
\begin{align}
\label{eq:gasp}
g_S^q = \theta_{\text{eff}}\,\frac{m_q }{f_a}\>, \qquad g_P^q = \frac{m_q }{f_a}\>.
\end{align}

Experimental results constrain the hypothetical axion mass severely. Observations of the neutrino signal from supernova 1987A and star cooling limits the axion mass to $\lesssim10$ meV~\cite{Raffelt:1999tx}. Correlated to small masses are induced long-range forces, which are in principle observable in laboratory experiments~\cite{Moody:1984ba}. Axions having larger masses are ruled out on the grounds that they would have produced already observable effects in astrophysical objects or terrestrial experiments. The situation for lighter axions is more optimistic because they not only avoid any conflict with current experimental data but are even strongly motivated on theoretical grounds~\cite{Freivogel:2008qc, Linde:1987bx}. Efforts in the experimental search for axions increased enormously when it was realized that they might provide a possible origin for dark matter~\cite{Bertone:2010zza, Raffelt:1999tx, Duffy:2009ig, Graham:2015ouw}.
 
Numerous other hypothetical light bosons have been proposed, which are close relatives of the axion and are collectively termed axionlike particles (ALPs). The situation for a generic scalar $\phi$ not related to the Peccei-Quinn mechanism, is in sharp contrast to axions. While the effective potentials in \eqnsref{eq:scalar-scalar} and (\ref{eq:pseudoscalar-scalar}) still hold, $g_S$ and $g_P$ are in this case a priori unrestricted free parameters with no relation to the breaking scale $f_a$, quark masses $m_q$, and effective minimum value $\theta_{\text{eff}}$. To the class of ALPs belong familons, the “pseudo-Nambu-Goldstone” bosons related to the spontaneous breaking of flavor symmetry~\cite{Davidson:1981zd, Wilczek:1982rv, Gelmini:1982zz}, majorons, which were proposed in order to understand neutrino masses~\cite{Chikashige:1980ui, Gelmini:1980re} and arions, which are the bosons related to the spontaneous breaking of chiral lepton symmetry~\cite{Anselm:1982}. In addition, hypothetical spin-0 or spin-1 gravitons have been hypothesized~\cite{Scherk:1979aj, Neville:1980dd, Neville:1981ut, Carroll:1994dq}. ALPs also arise in string theory as excitations of quantum fields extending into compactified extra space-time dimensions~\cite{Bailin:1987jd, Svrcek:2006yi}. The so-called axiverse is still another proposition in the context of string theory, which proposes the existence of many ultralight ALPs~\cite{Arvanitaki:2009fg}. Axions and ALPs have also been considered as a possible resolution to the hierarchy problem, addressing the question of why the Higgs mass is so much lighter than the Planck mass, contrary to expectations~\cite{Graham:2015cka}.
A natural question concerns the discrimination of axions from generic ALPs. Comparing the experimental results of nuclear EDM searches with those for new spin-dependent forces gives means to accomplish just that~\cite{Mantry:2014zsa, Mantry:2014} (also see Reference~\cite{Gharibnejad:2014kda}). 

In addition to spin-0 bosons, like axions and ALPs, many spin-1 particles have been hypothesized, which are related to new $U(1)$ gauge symmetries. In string theory and other extensions of the SM unbroken $U(1)$, gauge symmetries arise rather naturally~\cite{Cvetic:1995rj}. Corresponding to these are massless bosons, generically coined paraphotons~\cite{Holdom:1985ag}, while a prominent related proposal are dark photons~\cite{Ackerman:mha}. Furthermore, $Z'$ bosons related to the $Z$ boson of the SM arise in numerous theoretical models with broadly varying theoretically motivated masses and couplings to quarks and leptons~\cite{Langacker:2008yv}. The interaction of a massive spin-1 boson $Z'$ to a fermion $\psi$ is described by the Lagrangian
\begin{align}
\mathcal{L}_{Z'}=Z'_\mu\sum\limits_\psi \bar{\psi}\gamma^\mu\left(g_{V,\psi}+\gamma^5g_{A,\psi}\right)\psi\>.\label{eq:grav:lagr_spin1}
\end{align} 

The corresponding non-relativistic effective potentials are given by~\cite{Fadeev:2018rfl}
\end{paracol}
\nointerlineskip
\begin{align}
V_{VV}(r) &= \hbar c\,g_{V,1}g_{V,2}\,\frac{e^{-r/\lambda}}{4\pi r} + \delta_{VV} V(\bm s_1,\bm s_2)\,\ &&\text{vector-vector},\label{eq:vector-vector}\\
V_{AV}(r)&=\frac{g_{A,1}g_{V,2}}{4\pi}\,\bm{s}_1\!\cdot\!\left\{\!\frac{\bm{p}_1}{m_1}\!-\!\frac{\bm{p}_2}{m_2},\frac{\re^{-r/\lambda}}{r}\!\right\}\! +\! \delta_{AV} V(\bm s_1,\bm s_2) + (1\!\leftrightarrow\!2) &&\text{axial vector-vector},\label{eq:axialvector-vector}
\end{align}
\begin{paracol}{2}
\switchcolumn

\noindent where $\bm{p}_i$ are the configuration space momentum operators, and $\{,\}$ denotes the anti-commutator.
The operator $\bm{p}_i/m$ may by replaced by the classical velocity vector $\bm{v}_i$ for macroscopic masses but not for atomic or sub-atomic particles~\cite{Fadeev:2018rfl}. The terms $\delta_{VV} V(\bm s_1,\bm s_2)$ and $\delta_{AV} V(\bm s_1,\bm s_2)$ give non-vanishing contributions only if both interacting fermions are polarized (for the explicit expressions, see Reference~\cite{Fadeev:2018rfl}). For the \cannex{} setup, this would demand a spin-polarization in layers of both plates. Due to the strong magnetic background in this case, such a modification of the experimental setup is currently not envisaged. Therefore, $\delta_{VV} V(\bm s_1,\bm s_2)$ and $\delta_{AV} V(\bm s_1,\bm s_2)$ can be neglected for \cannex{}. Further remaining effective potentials can be found in Reference~\cite{Fadeev:2018rfl}, but, since they are not of immediate interest for \cannex{}, we refrain from enlisting them here.

\subsection{Dark Energy}
\label{sec:ODSIDE}
Concerning the mass/energy content of the present Universe, the largest fraction of the `dark side' is due to DE. As for DM, the origin of DE is unknown. However, since DM acts effectively at far smaller distance scales, our present state of knowledge on DM vastly exceeds the one on the comparatively elusive DE. 
As discussed in \secref{sec:vacuum}, the description of our Universe in terms of GR with an additional cosmological constant leads to a severe fine-tuning problem. Due to the rigidity of GR's theoretical framework concerning modifications altering the theory's behavior at large distance scales, it appears natural to introduce new additional degrees of freedom. Most commonly, the latter are realized in terms of a scalar field theory in order to maintain isotropy in the presence of non-vanishing field values. Those scalar fields used for the description of DE have been termed quintessence (for reviews, see Reference~\cite{Padmanabhan:2002ji, Peebles:2002gy, Frieman:2008sn, Linder:2007wa, Tsujikawa:2013fta, Joyce:2014kja}). Again, their existence induces new interactions, so-called fifth forces. However, precision astronomy within the solar system, as well as laboratory measurements, set tight limits on new interactions~\cite{Will:2014kxa,Joyce:2014kja}. Hence, these hypothetical new fields typically require some kind of `screening mechanism' to avoid conflicts with experimental limits. These mechanisms either suppress the scalar fields or their interaction with SM matter in regions of high density (i.e., the regions where strong constraints arise from observation), while they let the scalar prevail in the vast interstellar voids, where it may effectively drive the cosmic expansion. Among the different realizations of `screening mechanisms', one class employs non-linear self-interaction terms in the effective potential that induce a dependence of the effective potential and, thus, the field profile, on the energy density of the local environment. This class is further divided into three sub-classes depending on which way the scalar field is suppressed in dense environments~\cite{Burrage:2017qrf}. In the first sub-class, the effective mass associated with the field fluctuations becomes larger and, thus, the interaction range very short, while, in the second sub-class, the coupling of the field to matter becomes small. Finally, in the third sub-class, not all 
the mass of dense environments acts as a source of the field.
Specific models may belong to several sub-classes, i.e., the so-called `chameleon' models belong to the first and third sub-class~\cite{Khoury:2003rn, Khoury:2003aq, Mota:2006ed, Mota:2006fz, Waterhouse:2006wv}. This particular model has been intensively investigated and searched for in several laboratory experiments. A more recent model belonging mainly to the second sub-class is known as the `symmetron' model, which utilizes an effective potential similar to the Higgs potential~\cite{Hinterbichler:2010es, Hinterbichler:2011ca, Pietroni:2005pv, Olive:2007aj} (about its possible cosmological implications for small masses in the cosmological vacuum, e.g., of order of $10^3 H_0$, where $H_0$ is the current Hubble rate; see Reference~\cite{Hinterbichler:2011ca}).
At low energies, the effective Lagrangians of these two models are given by~\cite{Joyce:2014kja} 
\begin{align}
 \mathcal{L}_\text{Cha}&=\inv{2}\,\partial_\mu\phi\,\partial^\mu\phi - \frac{\Lambda^{4+n}}{\phi^n} - \rho\,\frac{\beta\,\phi}{M_\text{Pl}}\>, \\
 \mathcal{L}_\text{Sym}&=\inv{2}\,\partial_\mu\phi\,\partial^\mu\phi - \inv{2}\left(\frac{\rho}{M^2} - \mu^2\right)\phi^2 - \frac{\lambda}{4}\,\phi^4\>,
\end{align}
where $\phi$ denotes the respective scalar field. In each case, the Lagrangian depends explicitly on the mass density of the environment $\rho$ with corresponding coupling constants $\beta$ and $M$, respectively. Self-interactions depend on the couplings $\Lambda$ and $\lambda$ for chameleons and symmetrons, respectively. In addition, the latter depends also on a third parameter $\mu$, while $M_\text{Pl} = 1/\sqrt{8\pi G_N}$ appearing in the former is the reduced Planck mass. In both cases, the coupling to the mass density of the environment $\rho$ leads to small or vanishing field values for high mass densities, such as, e.g., inside the experimental setup. However, in a vacuum, the field takes on comparably large field values. For symmetrons, the effective potential is such that it acquires a nonzero vacuum expectation value (VEV) in low-density regions, where the field spontaneously breaks symmetry, couples to matter, and mediates a fifth force. However, in high density regions, the symmetry is restored, the field disappears, and is, thus, rendered invisible to any observation or measurement. In addition, of particular interest is the `dilaton' model, which belongs, again, mainly to the second sub-class. Its appeal draws from its string theory motivation, a context in which it was predicted by Damour and Polyakov~\cite{Damour:1994zq} (also see Reference~\cite{Brax:2010gi, Gasperini:2001pc}). In combination with an environment-dependent Damour-Polyakov coupling, this leads to a well-motivated screened DE model.
A weak point of these screened scalar field approaches to DE is that each of them still requires a certain amount of fine-tuning and also the presence of a non-zero cosmological constant. However, they bear some interest since they provide the simplest extension of the cosmological standard model $\Lambda$-CDM~\cite{Joyce:2014kja}.

Another class of `screening mechanism' relies on non-linearities in the kinetic sector. Screening is achieved by the so-called Vainshtein mechanism~\cite{Vainshtein:1972sx, Babichev:2013usa, Joyce:2014kja}. While Vainshtein-screened models are of great theoretical interest, they cannot be probed by laboratory experiments, in general \cite{Bellazzini:2017fep}. On the other hand, in Ref.~\cite{ArkaniHamed:2003uy}, DE is proposed to be represented by a ghost condensate, which is a constant-velocity scalar field. This fluid fills the Universe and mimics a cosmological constant due to its negative kinetic energy term. Its interaction with SM matter induces observable effects, like, e.g., Lorentz-violating effects, as well as new long-range spin-dependent interactions~\mbox{\cite{ArkaniHamed:2003uy, ArkaniHamed:2004ar}}. A completely different approach unrelated to quintessence, which does not rely on additional fields is the so-called modified Newtonian dynamics (MOND)~\cite{Milgrom:1983ca}. In this approach, Newtonian dynamics itself is modified for very low accelerations.
Another more exotic approach considers gravity as an emergent phenomenon rather than a fundamental interaction. It is, thus, termed `entropic-' or `emergent gravity', with its roots dating back to the 1970s. A recent version is due to Erik Verlinde, who proposed a conceptual model that describes gravity as a consequence of the `information associated with the positions of material bodies', thereby combining the thermodynamic approach to gravity with 't Hooft's holographic principle \cite{Verlinde:2010hp,Verlinde:2016toy}. 
Still another recent proposal along very different lines has given a possible quantum-optical explanation for DE in terms of a Lifshitz theory of the cosmological constant~\cite{Leonhardt:2019epj}, thus relating DE to the Casimir effect.
%
\section{The Quantum Vacuum and the Casimir Effect}
\label{sec:vacuum}
The expansion of the Universe was discovered by Hubble in 1929~\cite{Hubble:1929}. Much later, after studying type Ia supernovae data, this expansion was actually found to be accelerating~\cite{Perlmutter:1997zf, Riess:1998cb, Schmidt:1998ys}, which stands in stark contrast to long-held beliefs. The standard theory to describe gravity, as well as the dynamical evolution of our Universe as a whole, is GR. In its original formulation, however, GR predicts a decelerating Universe. In order to account for the accelerated expansion, the theory is amended by the so-called cosmological constant $\Lambda$. This extension of the theory was introduced by Einstein in 1917~\cite{Einstein:1917} but rejected by him in later years, as it was purely empirical. While $\Lambda$, thus, enables GR to describe accelerated cosmic expansion, it leads to a severe fine-tuning problem~\cite{Sola:2013gha}. This is due to the fact that, in addition to Einstein's original (bare) cosmological constant, additional contributions come from the zero-point energies of all quantum fields, i.e., those of the SM, as well as possible other ones that are still unknown. Even more, the Higgs potential also leads to an effective contribution during its phase transition related to electroweak symmetry breaking \cite{MARTIN2012566}.
While the measured value $\Lambda_\text{DE}\sim10^{-52}\,{\rm m}^{-2}$ necessarily contains all of these contributions, the zero-point energies of each particle's quantum field provides an infinite contribution to $\Lambda$, which is ordinarily subtracted via operator normal ordering. Keeping instead those contributions and rendering them finite by introducing momentum cutoffs at the electroweak scale, which provides the current upper limit of experimentally accessible energies, yields values for $\Lambda$ that are 55 orders of magnitude above $\Lambda_\text{DE}$ \cite{Sola:2013gha}. This `miraculous' reduction by many orders of magnitude to the small but non-zero value of $\Lambda_\text{DE}$ after summation of all individual contributions constitutes the \emph{cosmological constant problem}. Eventually, the latter results in a fine tuning problem, similar to the strong $CP$ problem, and gives a clear indication that our physical understanding is missing an important piece of the puzzle.

Since all astronomical observations rely on GR for their interpretation, it seems natural to assume that this theory needs to be modified. While it turns out to be simple to modify GR at short distance scales by including higher derivative terms, approaches that alter the theory's long-range behavior typically lead to theoretical inconsistencies. For this reason, most approaches amend GR by adding hypothetical new scalar fields instead of modifying it. Thereby, the cosmological constant is effectively promoted to the potential of those dynamical scalar fields. Such additional field degrees of freedom can naturally be identified with DE, as discussed in \secref{sec:ODSIDE}. 

DE with the correlated accelerated expansion of the Universe is but one example in which vacuum energy leads to measurable effects. More `down-to-earth' examples include spontaneous emission~\cite{Milonni:1975epj}, the Casimir effect, and the Lamb shift (for an excellent review, see Reference~\cite{Milonni:1994}). While the latter was measured already in 1947 in the famous Lamb-Retherford experiment~\cite{Lamb:1947zz}, the Casimir effect, theorized in 1948 by Hendrik Casimir~\cite{Casimir:1948}, had to wait half a century for a quantitative confirmation~\cite{Lamoreaux:1996wh}\footnote{Earlier measurements by Spaarnay~\cite{Sparnaay:1958wg}, as well as Blokland and Overbeek~\cite{Blokland:1978}, suffered from electrostatic and other interfacial disturbances.}. After the first modern measurements, a rapidly expanding field of Casimir research developed, investigating a wide variety of aspects. A non-exhaustive list contains the non-trivial dependence on spectral dielectric properties~\cite{Chen:2006zz,Torricelli:2011,Banishev:2012bh,Banishev:2013}, vanishing influence of layers much thinner than the electromagnetic penetration depth~\cite{Lisanti:2005}, non-trivial geometry~\cite{Tang:2017,Garrett:2018}, the existence of lateral forces~\cite{Chen:2002zzb}, and repulsive configurations~\cite{Munday:2009fgb}\footnote{It must be noted that similar results, though mostly ignored by the Casimir community, have been obtained earlier in colloidal science, with respect to the van der Waals force being a manifestation of the same quantum effects that give rise to Casimir forces.} (for a review of experimental data, see References~\cite{Iannuzzi:2015,Woods:2015pla}). Further motivation came from the significance of the Casimir effect for micro-electromechanical devices~\cite{Serry:1998,Ardito:2016,Broer:2013bxa,Tajik:2018}. However, a more fundamental question soon emerged that refers to the nature of virtual photons and their treatment in theory.

The commonly accepted description of (both thermal and zero-point) fluctuation forces between flat objects made of arbitrary materials is given by Lifshitz theory first presented 1956~\cite{Lifshitz:1956} and re-derived later using independent approaches~\cite{Dzyaloshinskii:1961,Ninham:1970,Genet:2003zz}. In this framework, the Casimir energy between two half-spaces separated by a distance $a$ at zero temperature and, hence, in the absence of thermal photons, is given by (for details of the derivation, see Reference~\cite{Bordag:2014})\\*[-3pt]
\end{paracol}
\nointerlineskip
\begin{align}
  E(a,T=0)=\frac{\hbar}{4\pi^2}\int\limits_0^\infty\!k_\bot\,{\rm d}k_\bot\int\limits_0^\infty\!{\rm d}\xi\left\{\ln\left[1-r^{(1,2)}_{TM}r^{(2,3)}_{TM}\re^{-2\kappa^{(0)} a}\right]+\ln\left[1-r^{(1,2)}_{TE}r^{(2,3)}_{TE}\re^{-2\kappa^{(0)}a}\right]\right\}\>,\label{eq:lifshitzT0}
\end{align}\mbox{}
\begin{paracol}{2}
\switchcolumn
\noindent with the (Fresnel) reflection coefficients\footnote{The real and imaginary parts of the complex dielectric susceptibilities are usually denoted as $\varepsilon(\omega)=\varepsilon'(\omega)+\ri \varepsilon''(\omega)$ with a similar notation for the complex magnetic susceptibilities.}
\begin{align}
\label{eq:Fresnel}
\begin{split}
 r^{(m,m')}_{TM}&=\frac{\vare^{(m)}(\ri\xi)\kappa^{(m')}(k_\bot,\ri\xi)-\vare^{(m')}(\ri\xi)\kappa^{(m)}(k_\bot,\ri\xi)}{\vare^{(m)}(\ri\xi)\kappa^{(m')}(k_\bot,\ri\xi)+\vare^{(m')}(\ri\xi)\kappa^{(m)}(k_\bot,\ri\xi)}\>, \\
 r^{(m,m')}_{TE}&=\frac{\mu^{(m)}(\ri\xi)\kappa^{(m')}(k_\bot,\ri\xi)-\mu^{(m')}(\ri\xi)\kappa^{(m)}(k_\bot,\ri\xi)}{\mu^{(m)}(\ri\xi)\kappa^{(m')}(k_\bot,\ri\xi)+\mu^{(m')}(\ri\xi)\kappa^{(m)}(k_\bot,\ri\xi)}\>,
\end{split}
\end{align}
depending on the complex dielectric and magnetic susceptibilities $\vare^{(m)}(\ri\xi)$ and $\mu^{(m)}(\ri\xi)$, respectively, of the materials indexed by $m=0$ for the gap and $m=\{1,2\}$ for the two half-spaces. Furthermore, we have 
\begin{align}
 \kappa^{(m)}(k_\bot,\ri\xi)=\sqrt{k_\bot^2+\vare^{(m)}(\ri\xi)\xi^2/c^2}\>,
\end{align}
being a function of the transversal momentum $k_\bot$ (parallel to the interfaces), and complex frequency $\ri\xi$ related to the quantized vertical direction. The fluctuation-dissipation theorem states that the occurrence of fluctuations implies the presence of dissipation, and vice versa. This should be valid for both virtual and thermal fluctuations, as is confirmed by, for example, the non-zero ohmic resistance of metal-like conductors at $T=0$. Accordingly, it was assumed also in the derivation of \eqnref{eq:lifshitzT0} that the fluctuation-dissipation theorem holds true and that the dielectric response of the materials to vacuum fluctuations can be described by the functions $\vare(\omega)$ and $\mu(\omega)$. These can actually be measured from the far infrared to x-ray frequencies using spectroscopy, and at low frequencies using electrical measurements. As $\mu$ does not differ much from unity for most materials, it is usually ignored by setting $\mu^{(m)}\to 1$. In \eqnref{eq:lifshitzT0}, the dielectric functions appear with complex arguments\footnote{The appearance of complex functions results from the infinite integration of the mode density functions over real frequencies. As the latter have poles, the integration contour splits into a half-circle at infinity and an integration from $-\ri\omega$ to + $\ri\omega$. The former term vanishes due to ${\rm Im}[\vare(\omega)]\to0$ for $|\omega|\to\infty$, leaving the integration over the complex frequency axis.}. The functions $\vare(\ri\xi)$ (and $\mu(\ri\xi)$) are inaccessible to direct measurements and have to be obtained from data on $\vare(\omega)$ via Kramers-Kronig transforms~\cite{Landau:2013}

\begin{align}
 \vare(\ri\xi)=1+\frac{2}{\pi}\,{\rm P}\!\!\int\limits_0^\infty\!{\rm d}\omega\,\frac{\vare''(\omega)\xi}{\omega^2+\xi^2}+\Pi(\xi)\>,\label{eq:kk-relation}
\end{align}

\noindent where the principle-value integral depends exclusively on the \emph{complex part} $\vare''(\omega)$ of the dielectric function. The term $\Pi(\xi)=0$ if $\vare''(\omega)$ has a first-order pole at $\omega=0$, which is the case if dissipative materials are described by a Drude model (see below), while $\Pi(\xi)=\omega_p^2/\xi^2$ for $\vare''(\omega)$ having a second-order pole at $\omega=0$, as is the case for an electron gas described by the dissipationless plasma model. $\omega_p=\sqrt{\rho e^2/\vare_0 m^*}$ denotes the plasma frequency for charge carrier density $\rho$ and effective electron mass $m^*$ in the material. In order to execute the integral in \eqnref{eq:kk-relation}, extrapolation of experimental data on $\vare(\omega)$ (and $\mu(\omega)$) to zero frequency is required.

From the Maxwell equations, it follows~\cite{Landau:2013} that, for $\omega\to 0$, in a material with finite conductivity $\sigma$, we have $\vare(\omega)\to\ri 4\pi\sigma/\omega$. This concept is captured by the Drude model 

\begin{align}
 \vare_D(\omega)=1-\frac{\omega_p^2}{\omega(\omega+\ri\gamma)}+\vare_c\>,\quad \text{with}\quad\vare_c=\sum\limits_i\frac{f_i}{\omega_i^2-\omega^2-\ri\gamma_i\omega}\>,\label{eq:epsilon_model}
\end{align}

\noindent where the contribution $\vare_c$ of core electrons is written as a sum over oscillators with amplitudes $f_i$, resonance frequencies $\omega_i$, and damping $\gamma_i$. The relation between the free electron contribution and finite DC conductivity $\sigma$ is given by $\sigma=\omega_p^2/4\pi\gamma$~\cite{Ashcroft:1976}, leading to exactly the low frequency limit of $\vare(\omega)$ stated above that can actually be measured. The Drude model is valid in a wide range of frequencies and agrees with physical observation. On the other hand, the plasma model $\vare_P=1-\omega_p^2/\omega^2+\vare_c$ assumes that electrons can respond freely and elastically to electromagnetic fields, which is true for the electron gas of an ideal metal at high frequency. We should be aware at all times that neither $\vare_D$ nor $\vare_P$ are rigorous models of the dielectric response of materials but should be considered as approximations, serving only to discuss the issue of dissipation here.

In order to compare experiment and theory, one more ingredient is required. In fact, the presence of boundary conditions not only leads to quantization of the spectrum of vacuum fluctuations, but also of the thermal radiation field of the involved bodies at $T\neq0$. The important point here is that, in the derivation of Lifshitz theory, it is assumed that the boundary conditions are described by the functions $\varepsilon(\omega)$ and $\mu(\omega)$ for \emph{both} fields---vacuum and thermal fluctuations. On this basis, the fluctuation-dissipation theorem is applied. Formally, the total energy containing \emph{both} the vacuum and thermal radiation field energies can be obtained by the replacement~\cite{Bordag:2014} $\frac{\hbar}{2\pi}\int_0^\infty \!{\rm d}\xi\ \leftrightarrow\ k_B T{\sum_{n=0}^\infty}'$ in \eqnref{eq:lifshitzT0}, where we use the standard SI nomenclature for natural constants. The sum on the right side runs over discrete \emph{Matsubara} frequencies $\xi_n=2\pi k_B T n/\hbar$, which are just the poles of the partition function for the thermal fluctuations. In the sum, the prime indicates that the $n=0$ term receives an additional factor $1/2$.

As discussed above, the procedure using the Drude model to extrapolate data on $\vare''(\omega)$ to $\omega=0$ to obtain $\vare(\ri\xi)$ that, in turn, is inserted into the $T\neq 0$-version of \eqnref{eq:lifshitzT0} is in agreement with Lifshitz theory and observation. However, surprisingly, it leads to results that are incompatible with experimental data on the Casimir force~\cite{Bimonte:2016,Liu:2019xgg}, and was furthermore shown to violate the third law of thermodynamics, i.e., that the entropy has to vanish for zero temperature~\cite{Bezerra:2002}. The discrepancy between experimental data and the theoretical prediction was noted already 20 years ago, although the first modern experiment initially reported agreement with the Drude procedure~\cite{Lamoreaux:1996wh} (which was revised later~\cite{Lamoreaux:2010}). In the community, the reactions to this problem were mixed, leading to decade-long discussions and investigations of the subject. A detailed review can be found in Ref.~\cite{Mostepanenko:2021cbf}. On the experimental side, electrostatic patches, roughness, distance determination, and other problems were excluded to be the cause of the discrepancy. High-accuracy data for the Casimir interaction between metals are now available from two different groups~\cite{Bimonte:2016,Bimonte:2021,Liu:2019xgg}, both of which exclude the Drude model---and henceforth seemingly dissipation---with high confidence at short separations. It is worth emphasizing that, naturally, there is no dissipation for virtual photons at $\omega\neq0$ as this would correspond to energy generation from the vacuum. Instead, the question focuses on dissipation at $\ri\omega=\xi_0=0\ri$ (at the $n=0$ Matsubara frequency), which is not subjected to this problem. As indicated above, the $\omega=0$ term is related to DC resistivity, which is known to exist for all non-superconducting materials, for which researchers are puzzled by the experimental results. It might be speculated that the incompatibility between theory and experiment comes from the fact that both the Drude and the plasma models are simple approximations only. In this respect, graphene is interesting, as the dielectric response for this material given via the polarization tensor can be derived from first principles using the Dirac model~\cite{Bordag:2009fz,Fialkovsky:2011pu}. This description does include dissipation and leads to an expression for the entropy that is compatible with thermodynamics (see, e.g., Reference~\cite{Klimchitskaya:2020qqk}). In the only experiment probing the Casimir force between a gold sphere and a suspended graphene sheet~\cite{Banishev:2013}, good agreement with the Dirac model was found~\cite{Klimchitskaya:2014axa}.

For semiconductors, yet another unexpected finding was made. If the charge carrier density $\r$ is below the critical density $\r_\text{cr}$ for the insulator/metal (Mott-Anderson) phase transition, the material is in the dielectric (insulating) state, and the contribution of free charge carriers in \eqnref{eq:epsilon_model} has to be neglected entirely. If, on the other hand, $\r>\r_\text{cr}$, this contribution needs to be considered~\cite{Klimchitskaya:2012fn}. This effect was demonstrated for silicon surfaces with different $\r$~\cite{Chen:2006zz}, a silicon substrate switched from $\r<\r_\text{cr}$ to $\r>\r_\text{cr}$ by light~\cite{Chen:2007kqw}, and an ITO substrate~\cite{Chang:2011} in metallic and dielectric state. Unfortunately, all experiments performed to date probing semiconductor test bodies were of insufficient accuracy to distinguish between Drude and plasma model for the metallic state. However, proposals exist to achieve just that~\cite{Bimonte:2019ahv}. Again, and similarly as for metals, if the DC conductivity of insulators is taken into account, then the Casimir entropy does not vanish in the zero-temperature limit~\cite{Geyer:2005ap}, thereby violating the third law of thermodynamics.

Experimental data, thus, indicates that dissipation of conduction band electrons has to be neglected at short separation up to $\SI{4}{\micro\metre}$~\cite{Bimonte:2021}. In this range, the major contribution to the Casimir energy comes from virtual photons. In fact, the contribution of \emph{thermal} fluctuations is small at separations smaller than the thermal wavelength $\lambda_T=\hbar c/k_B T$, which is \SI{7.6}{\micro\metre} at room temperature\footnote{Note that, for graphene, the thermal contribution to the Casimir effect is significant at much smaller separations~\cite{Gomez-Santos:2009rxl}. However, this effect could not be measured yet.}. Therefore, the present experiments demonstrating agreement with the plasma model have probed to a large extent the boundary conditions on virtual photons. 
Physical intuition tells us that thermal photons do carry energy and, thus, necessarily obey the fluctuation-dissipation theorem. As speculated earlier~\cite{Sedmik:2018kqt,Klimchitskaya:2019nzu,Liu:2019sny}, virtual (off-shell) and thermal (on-shell) photons, thus, might have to be treated differently in theory and might not be subjected by the same boundary conditions, unlike assumed in the standard derivation of Lifshitz theory~\cite{Bordag:2014}. A hint in this direction may be seen in the polarization tensor approach used to describe the dielectric functions of graphene~\cite{Klimchitskaya:2020qmy,Bordag:2009fz,Fialkovsky:2011pu}. 
Here, the material's response is non-local in the sense that it depends on both the frequency and the wave vector (in contrast to the dielectric functions for the Drude and plasma models that depend on (complex) frequencies only).
It was shown that the Casimir entropy based on the Dirac model approach satisfies the third law of thermodynamics~\cite{Klimchitskaya:2020qqk} \emph{and} describes the physically important effect of dissipation. Therefore, the `Casimir puzzle' (see the review in Reference~\cite{Mostepanenko:2021cbf}), referring to the problem of the seemingly incompatibility of the proven effect of dissipation with experiment and fundamental thermodynamics, does not exist for graphene. On the basis of these contemplations, a new approach for the description of dielectric functions of metals was proposed recently~\cite{Klimchitskaya:2020qmy}. Unlike for graphene, this description is not based on first principles but attempts to include non-locality in the simplest possible way. The resulting modified Drude-like dielectric functions read
\begin{align}
\label{eq:NLDrude}
    \vare_\text{D,NL}^T(\omega, k_\bot)&=1-\frac{\omega_p^2}{\omega(\omega+\ri\gamma)}\left(1+\ri\frac{\zeta^T v_F k_\bot}{\omega}\right)\,,\nonumber\\
    \vare_\text{D,NL}^L(\omega, k_\bot)&=1-\frac{\omega_p^2}{\omega(\omega+\ri\gamma)}\left(1+\ri\frac{\zeta^L v_F k_\bot}{\omega}\right)^{-1}\,,
\end{align}
where $v_F$ is the Fermi velocity, and $\zeta^{T,L}$ are free parameters of the model. Following Ref.~\cite{Klimchitskaya:2020qmy}, we choose them to be $\zeta^T=7$ and $\zeta^L=1$, but the values remain purely empiric to fit experimental data. 
A non-local description is only possible if the geometry is invariant under translation which, for parallel plates, is only true in the two directions parallel to the surface but not in the quantized orthogonal direction~\cite{Klimchitskaya:2007a}. For this reason, two functions, $\vare_\text{D,NL}^L(\omega, k_\bot)$ and $\vare_\text{D,NL}^TL(\omega, k_\bot)$, emerge for the longitudinal and transversal degrees of freedom, respectively, depending not on the three-dimensional wave vector $\mathbf{k}$ but only on the transversal (parallel to the plates) wave vector $k_\bot$, similar as for the case of graphene. 
The resulting reflection coefficients for non-local dielectric response were given for the case of equal (metal) material on either side of a gap `filled' by vacuum~\cite{Klimchitskaya:2020qmy}
\end{paracol}
\nointerlineskip
\begin{align}
\label{eq:reflection_coeff_NL}
   r^{(m,0)}_{TM,\text{NL}}&=\frac{\vare_\text{D,NL}^{(m),T}(\ri \xi, k_\bot)\kappa^{(0)}-\kappa^{(m),T}-k_\bot \left[\vare_\text{D,NL}^{(m),T}(\ri \xi, k_\bot)-\vare_\text{D,NL}^{(m),L}(\ri \xi, k_\bot)\right]/\vare_\text{D,NL}^{(m),L}(\ri \xi, k_\bot)}{\vare_\text{D,NL}^{(m),T}(\ri \xi, k_\bot)\kappa^{(0)}+\kappa^{(m),T}+k_\bot \left[\vare_\text{D,NL}^{(m),T}(\ri \xi, k_\bot)-\vare_\text{D,NL}^{(m),L}(\ri \xi, k_\bot)\right]/\vare_\text{D,NL}^{(m),L}(\ri \xi, k_\bot)}\>, \\
 r^{(m,0)}_{TE,\text{NL}}&=\frac{\mu^{(m)}(\ri\xi)\kappa^{(0)}-\kappa^{(m),T}}{\mu^{(m)}(\ri\xi)\kappa^{(0)}+\kappa^{(m),T}}\>,
\end{align}
\begin{paracol}{2}
\switchcolumn

\noindent where we have abbreviated $\kappa^{(0)}=\kappa^{(0)}(k_\bot,\ri\xi,\varepsilon^{(0)}=1)$ and $\kappa^{(m),T}=\kappa^{(m)}(k_\bot,\ri\xi,\varepsilon^{(m)}=\vare_\text{D,NL}^T(\omega, k_\bot))$. These coefficients can readily be used with the Lifshitz expression for the Casimir energy~\eqref{eq:lifshitzT0} and pressure~\eqref{eq:Peq} (see \secref{sec:prospects}).
For thermal on-shell photons, the dispersion relation $k_\bot^2=\epsilon \omega^2/c^2$ holds for $|\epsilon|\geq1$, representing the effective response of the material that might depend on $\omega$ and $\mathbf{k}$. Hence, the additional terms proportional to $v_F$ in \eqnref{eq:NLDrude} are of order $v_F/c$. As for metals, $v_F\lesssim c/300$~\cite{Gall:2016}, these additional terms are small for on-shell photons for all $k_\bot$. For off-shell photons, however, $k_\bot^2\neq\epsilon \omega^2/c^2$, and the additional term may take arbitrarily large values. Note, however, that, for all $\zeta^T\neq 0$, $\vare_\text{D,NL}^T(\omega, k_\bot)$ bears a plasma model-like second-order pole and corresponding reflection coefficients for $\omega\to0$. Therefore, the functions in \eqnref{eq:NLDrude} implement dissipation together with non-locality. It was shown that the experimental data can be brought into agreement with Lifshitz theory, \emph{including dissipation}, if the non-local reflection coefficients in \eqnref{eq:reflection_coeff_NL} are used in the calculation of the Casimir pressure. We remark, however, that this success does not yet state a solution to the Casimir puzzle, as the non-local Drude model is purely empirical. However, the model gives a hint that the differences between virtual and thermal photons may have to be investigated further with respect to how photons on and off the mass shell interact with matter.

The success of the non-local Drude model renews the need for high-accuracy experimental data to distinguish between not only the Drude and plasma models at all separations but also between the plasma model and the non-local Drude model.
A definite answer could be given in several ways. The most obvious approach is to measure at larger separations with $a>\lambda_T$. However, the forces there are too weak to perform precision measurements with most present experiments. In fact, a recent analysis~\cite{Bimonte:2021} of experimental data concluded that the Drude model can be ruled out for $a<\SI{4.8}{\micro\metre}$, while at larger separation the situation remains unclear. The second way is to measure at higher temperature, which seems even more challenging, as this necessarily also increases systematic disturbances in sensitive experiments. A third and more general way to investigate thermal effects is to use 2D-materials, such as graphene sheets, for which these effects are significantly stronger at small separations~\cite{Bimonte:2019ahv}, and the dielectric response can be derived from first principles.

Experiments are notoriously difficult, as the force $F_C$ sharply falls off with object separation $a$, for which measurements are performed typically at $d< \SI{1}{\micro\metre}$ using atomic force microscopes~\cite{Mohideen:1998iz,deMan:2009zz,Torricelli:2010a,Garrett:2018,Liu:2019xgg}, or micro-machined torsion balances~\cite{Decca:2003td,Iannuzzi:2004,Chan:2011za,Decca:2016zmo}. At small separations, surface roughness~\cite{Klimchitskaya:1999gg,Neto:2006,vanZwol:2008,Broer:2012fc,Sedmik:2013vna} and local variations of the Fermi potential (patches)~\cite{Speake:2003zz,Kim:2009mr,Behunin:2012zr,Garrett:2020acf} have to be considered. Another problem is that the benefit of circumventing the issue of measuring and maintaining parallelism by using a plate versus sphere geometry, is bought at the expense of the resulting Casimir force being generated only in a minor fraction of the total curved surface. Due to the latter fact, the force amplitude (and sensitivity) are decreased, and the dependence on omnipresent local corrugations is amplified~\cite{Bezerra:2011,Sedmik:2013vna}. These problems are only slightly alleviated at larger distance (up to $\sim$$\SI{10}{\micro\metre}$), where macroscopic torsion balances~\cite{Masuda:2009vu,Sushkov:2011md} with cm-sized curved test bodies are used. As will be shown below, the small effective force-generating area $A_\text{eff}\approx\pi Ra$~\cite{Sedmik:2013vna,vanZwol:2009} is indeed one of the main limitations in present experiments.
%
\section{Force Metrology with Macroscopic Objects}
\label{sec:pp}
 Continuous technological progress has led to the development of force metrology, which has produced some of the tightest limits on several hypothetical DM and DE models. Prototypical examples for such experiments are torsion pendula that have already been used by Cavendish to measure the gravitational constant. Today, torsion pendula still play an important role to find limits on the parameters of DE and DM models in the range \SI{10}{\micro\metre}--\SI{10}{\milli\metre}~\cite{Adelberger:2009zz,Tan:2020vpf,Zhao:2021anp}. At smaller separations, Casimir experiments using atomic force microscopes (AFM)~\cite{Liu:2019xgg}, or (micro-)torsional balances~\cite{Chen:2014oda,Bimonte:2016,Bimonte:2021}, are more sensitive. Intermediate scales around \SI{10}{\micro\metre}, however, are not covered well by experiments~\mbox{\cite{Chen:2014oda,Geraci:2008hb,Masuda:2009vu,Sushkov:2011md, Lee:2020zjt,Tan:2020vpf}}. However, in this range, interesting observations could be made. Firstly, as discussed above, in the thermal regime the thermal Casimir force could be measured unambiguously, thereby improving knowledge about the correct treatment of virtual (vacuum) fluctuations and thermal fluctuations. Secondly, distances $\gtrsim \SI{20}{\micro\metre}$ fall into the QCD `axion window' $\SI{10}{\milli\electronvolt}<m_A<\SI{1}{\micro\electronvolt}$~\cite{Raffelt:1999tx}, where the axion---if it existed---would be expected. In \tabref{tab:exp-comparison}, we compare several experiments that have measured either Casimir or non-Newtonian interactions in the range 1--\SI{10}{\micro\metre}. The figure of merit in terms of the ability to measure interfacial interactions at larger separation is the pressure sensitivity, which relates the absolute sensitivity to the effective available force-generating area $A_\text{eff}$. For spherical surfaces of curvature $R_c$, we have $A_\text{eff}\approx aR_c\pi$~\cite{vanZwol:2009,Sedmik:2013vna}, which depends on surface separation, while, for parallel plates, $A_\text{eff}=A$. 
 
 \begin{specialtable}[H] 
 \caption{Order-of-magnitude-comparison of several present experimental setups in terms of geometry, size of the force-generating area $A_\text{eff}$, and sensitivity to forces and pressures.\label{tab:exp-comparison}}
\begin{tabular*}{\hsize}{@{}@{\extracolsep{\fill}}cccccc@{}}
\toprule
\textbf{Experiment,}& \textbf{Object}& \textbf{Force Gen.} & \multicolumn{2}{c}{\textbf{Sensitivities}}\\
     \textbf{Geometry}&\textbf{Size}&\textbf{Area}& \multicolumn{1}{c}{\textbf{Force}} & \multicolumn{1}{c}{\textbf{Pressure}} & \textbf{Ref.}\\
     & &  \textbf{[cm\boldmath$^2$]}& \textbf{[pN]}&\textbf{[pN/cm\boldmath$^2$]}\\
     \hline
    \multicolumn{6}{l}{\emph{interfacial measurements}}\\
      \multicolumn{1}{l}{AFM-type, sphere/plane} & \SI{100}{\micro\metre} & $10^{-6}$ & $0.1\phantom{^-}$ & $10^5\phantom{^-}$ & \cite{Xu:2018}\\
      \multicolumn{1}{l}{torsional balance, sphere/sphere} & \SI{10}{\centi\metre} & $10^{-2}$ & $10\phantom{-5}$ & $10^3\phantom{^-}$ & \cite{Masuda:2009vu}\\
      \multicolumn{1}{l}{micro-oscillator, sphere/plane} & \SI{100}{\micro\metre} & $10^{-6}$& $10^{-4}$ & $10^{2}\phantom{^-}$ & \cite{Bimonte:2016}\\
     \multicolumn{6}{l}{\emph{prospective}}\\
      \multicolumn{1}{l}{\cannex{} oscillator, plane/plane} & \SI{1}{\centi\metre} & $\phantom{0}1\phantom{-5}$ & $0.1\phantom{-}$ & $0.1\phantom{-}$ & \cite{Klimchitskaya:2019fsm}\\
     \hline
     \multicolumn{6}{l}{\emph{Cavendish-type measurements}}\\
      \multicolumn{1}{l}{micro-cantilever, cube/plane}  & \SI{100}{\micro\metre} & $10^{-5}$ & $10^{-6}$ & $0.1\phantom{-}$ &\cite{Geraci:2008hb}\\
      \multicolumn{1}{l}{torsion balance, patterned plates} & \SI{10}{\centi\metre} & $10\phantom{-5}$ & $10^{-3}$ & $10^{-4}$&\cite{Lee:2020zjt}\\
\bottomrule
\end{tabular*}

\end{specialtable}

 For precision measurements of Casimir forces, torsion pendula allowed for the most sensitive measurements at their time. The large surface curvature radii $R_c$ of tens of centimeters allow decent effective force-generating areas for separations in the micrometer range. However, difficulties in producing sufficiently smooth curved surfaces~\cite{Bezerra:2011} and other technical issues rendered further progress challenging. Benefiting from their versatility, AFMs with spherical test objects are still used widely. Despite the typically small force generating surfaces and correspondingly low pressure sensitivity, these experiments still give competitive data~\cite{Liu:2019xgg} at small separations. While suffering from the same problem of small $A_\text{eff}$, micro-torsional oscillators~\cite{Decca:2003td,Iannuzzi:2004,Chan:2011za,Decca:2016zmo,Bimonte:2021} allow for higher sensitivity, and presently set the reference in terms of precision Casimir measurements. This is especially true for differential measurements. Elimination of background, electromagnetic, and Casimir forces by means of a shield between the interacting bodies allows for a further increase in sensitivity. This strategy is followed in Cavendish-type experiments. Naturally, such experiments can only probe differential gravity-like interactions not blocked by the shield. Therefore, the figure of merit to compare respective experiments would be the force sensitivity per volume. However, in order to compare Cavendish-type with interfacial measurements, we only give the sensitivity per area. Using a micro-machined gold test mass on a cantilever in cryogenic environment, aN sensitivity was achieved~\cite{Geraci:2008hb}. For a long time, the reference was set by the E\"otvos group experiments using a torsional pendulum with angular modulation techniques~\cite{Lee:2020zjt}\footnote{Note that the actual quantity measured in torsion pendula experiments is a torque. For the purpose of comparing the order-of-magnitude sensitivity of Ref.~\cite{Lee:2020zjt} to other experiments with force detection, we divide the torque sensitivity $0.01\,$fNm by the effective radius \SI{16.5}{\milli\metre} of the test mass. For comparing the pressure (again only as rough estimation), we assume the entire torque-generating area of \SI{10}{\centi\metre^2} to contribute homogeneously.}. Recently, an improved torsion balance setup provided slightly better sensitivity~\cite{Tan:2020vpf}. In comparison, the Casimir And Non-Newtonian force EXperiment (\cannex{}) uses a mesoscopic linear oscillator with large \SI{1}{\centi\metre^2} force-generating area. This latter fact is the reason, why \cannex{}, despite its comparably humble force sensitivity of $0.1\,$pN, could be able to reach a superior pressure sensitivity for an interfacial force measurement.
 In addition, from the theoretical point of view, this geometry is advantageous, as it closely resembles the one-dimensional approximation of two semi-spaces allowing for closed-form solutions for many problems. On the downside, symmetry demands accurate control over parallelism, which adds complexity, and results in increased sensitivity to large-scale surface deformations. Moreover, at surface separations of a few (tens of) micrometers, dust is a major issue~\cite{Bressi:2002fr,Antonini:2008bb,Sedmik:2018kqt,Lee:2020zjt}. For this reason, several earlier attempts to implement the parallel plate geometry in Casimir measurements have been~unsuccessful.
 
In \cannex{}, several of the technical issues were solved by design. However, the proof-of-principle~\cite{Sedmik:2018kqt} suffered from a parasitic AC signal coupling, and broken thermal control that could not be repaired at the time due to exclusive remote access to the setup. As analyzed later~\cite{Klimchitskaya:2019fsm}, the main error sources are vibrations, thermal variation, and electrostatic \emph{patch effects} (at small separations only). In \secref{sec:design}, we present a new design that significantly reduces all major error sources and shall enable measurements of \mbox{pressures $<0.1\,$pN/cm$^2$} between parallel plates separated by 3--\SI{30}{\micro\metre}.
%
\section{New Design}
\label{sec:design}
Based on a recent analysis of systematic errors~\cite{Klimchitskaya:2019fsm}, technical issues~\cite{Sedmik:2018kqt}, and future measurement options~\cite{Klimchitskaya:2019nzu}, we have developed a new design to improve the existing \cannex{} setup in several ways.
Firstly, we augmented insulation against vibrational and thermal disturbances significantly. Secondly, we improved the measurement sensitivity and reduced systematic errors by implementing all-optical measurement methods, additional alignment procedures, and active thermal control of the sensor element. The latter measure also allows now for measurements of Casimir interactions and heat transfer out of thermal equilibrium~\cite{Klimchitskaya:2019nzu}. In the following, we discuss the different systems and give an updated and comprehensive error estimation.

\subsection{Core Setup and Measurement Principle}
\label{sec:des:core}
The so-called \emph{core} contains the actual measurement setup, which is depicted schematically in \figref{fig:core_schematic}. This setup is placed inside a small (core) vacuum chamber that allows for pressures between high vacuum ($\sim$$10^{-9}\,$mbar) and $\sim$$1\,$mbar of an arbitrary gas. The core chamber is suspended on a seismic, acoustic, and thermal isolation system inside a larger vacuum chamber (see \secref{sec:des:vibr}). In this section, we describe the core only.
As a force sensor, \cannex{} uses a mesoscopic mass-spring system formed by three interleaved helical coils etched, together with the disk-shaped test mass (upper plate) from a single silicon wafer~\cite{Almasi:2015zpa}. As shown schematically in \figref{fig:core_schematic}, the test mass is placed parallel and co-centered above the flat end face of a polished glass cylinder (lower plate). Both plates are optically smooth with maximum deviation from perfect flatness of \SI{10}{\nano\metre} on all length scales (see \secref{sec:des:flat}) and coated with a \SI{1}{\micro\metre} thick layer of gold above a thin TiW adhesion layer. The sensor is mounted in a rigid frame that can be moved for initial alignment vertically and horizontally by \SI{5}{\milli\metre} with respect to the common base plate using drift-free `stick-slip' piezo translators. Above the sensor, a dome-shaped thermal shroud coated with graphite serves as contact-free thermal interface to the sensor (see \mbox{\secref{sec:des:thermal}}). Together with the surrounding structure, the shroud creates a Faraday cage around the sensor which has a single opening towards the lower plate. The latter is mounted on a rigid structure connected to three linear piezo-electric actuators (range \SI{150}{\micro\metre}) that allow to finely adjust the relative tilt and separation between the two plates. The plates can be brought to a maximum distance of \SI{5}{\milli\metre} using the stick-slip translator, which opens up a window on opposite sides of the electrostatic shielding of the core. In this position, the surfaces of both plates have a free line of sight towards an Ar ion gun mounted in the core chamber, as shown in the figure. Another addition with respect to the previous implementation are high-power UV LEDs with wavelength $\sim$$280\,$nm mounted in two rows around the plates.

\begin{figure}[H]
    \includegraphics{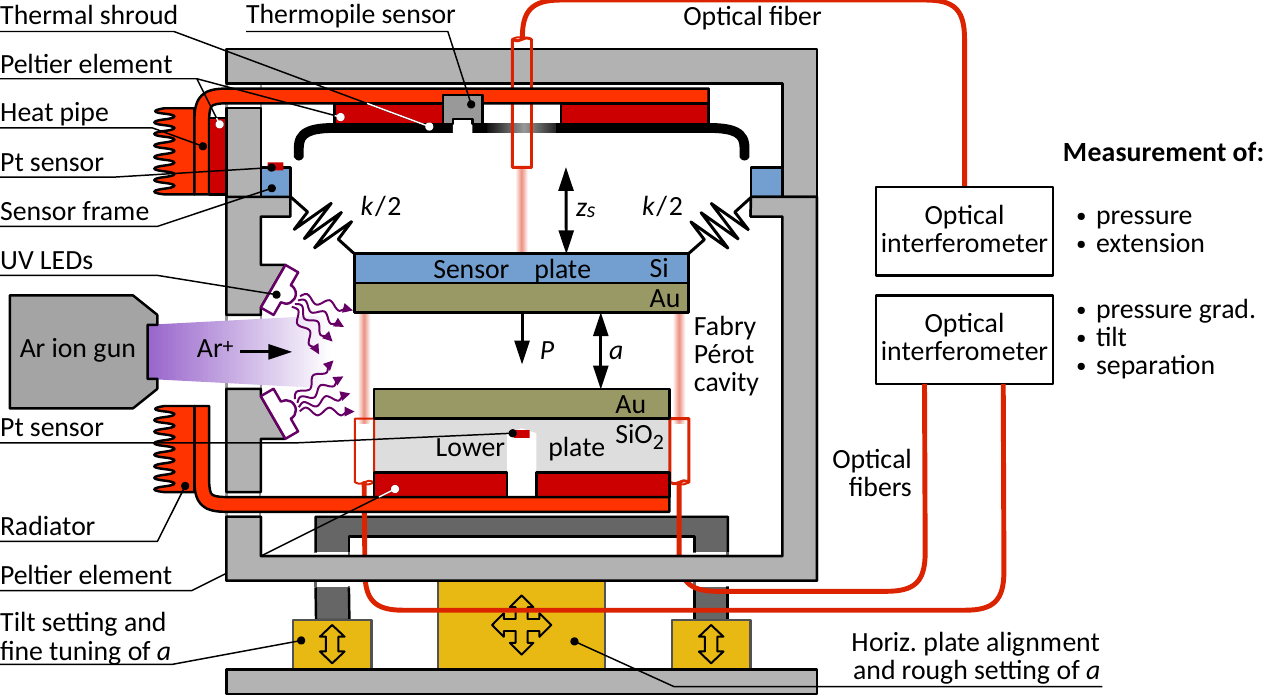}
    \caption{Simplified schematic representation of the \cannex{} core setup showing the fiber-optic detection mechanism, major mechanical components, and parts of the thermal control system. Note that the real setup has a $120^\circ$ symmetry in most parts. In this two-dimensional representation, a partial $180^\circ$ symmetry is depicted instead.\label{fig:core_schematic}}
\end{figure}
A pressure $P$ between the plates of area $A$ generates extensions $\Delta z$ of the springs having an effective elastic constant $k\approx 0.13\,$N/m~\cite{Sedmik:2018kqt}. These extensions are measured via a Fabry-P\'erot interferometer formed by the end of a cleaved optical fiber and the reflective upper surface of the sensor. This allows for a direct way to determine $P=k\Delta z/A$ from the interferometer data. We use active feedback with wavelength modulation to keep the interferometer signal at its quadrature value (zero transition of the interferometric fringe), independent of the extension $\Delta z<\SI{20}{\nano\metre}$ due to interfacial forces at all separations. 
Three additional Fabry-P\'erot fiber interferometers arranged symmetrically around the rim of the lower plate view the sensor from below. These have two different purposes. Firstly, they are used 
to track the sensor's resonance frequency shift 
\begin{align}
    \Delta \omega_0=\omega_0-\sqrt{\omega_0^2-(A/m_\text{eff})P'(a)}\>,\label{eq:freq_shift}
\end{align}
due to pressure gradients $P'(a)=\partial P(a)/\partial a$ using a phase-locked loop (PLL)~\cite{Chang:2012fh,Sedmik:2018kqt}. Here, $\omega_0=2\pi f_0=\sqrt{k/m_\text{eff}}$ is the mechanical resonance frequency, $m_\text{eff}$ is the effective sensor mass, and $A=\SI{1}{\centi\metre^2}$ is the force-generating area. In the experiment, the sensor is excited electrostatically at frequency $\omega_{ex}=(\omega_0-\Delta \omega_0)/2$ using a well-known excitation signal $v_{ex}(t)=V_{ex}\sin\omega_{ex} t$ with the amplitude $V_{ex}\approx \SI{1}{\milli\volt} (a/a_\text{ref})^2$ being adapted to assure an equal signal level at all separations ($a_\text{ref}=\SI{10}{\micro\metre}$). The actual excitation signal is regulated by a feedback circuit maintaining the signal level. With respect to frequency shift measurements, silicon with its low thermoelastic damping coefficient~\cite{Denu:2017} and large elastic modulus~\cite{Hopcroft:2010} is advantageous, as it offers a high mechanical quality factor $Q\sim 10^4$~\cite{Sedmik:2018kqt} and, thereby, a small detection bandwidth $f_\text{BW}\approx\SI{3}{\milli\hertz}$. On the other hand, the large Q-factor implies long integration times $\tau\sim (2\pi f_\text{BW})^{-1}$. In addition, ring-down of the sensor after seismic or operational disturbances takes a long time, as well, for which typical measuring times amount to roughly an hour per point. The second purpose of the three independent interferometers is to actively measure and control parallelism between the two plates, as discussed next. 

\subsection{Parallelism}
\label{sec:des:parallel}

Previously~\cite{Sedmik:2018kqt}, we used a capacitive modulation scheme to detect deviations from perfect parallelism. However, the combined problems of thermal drift and long integration times limited the performance to a few \si{\milli\radian} instead of the targeted \SI{1}{\micro\radian}. In addition, the excitation voltage of the capacitive bridge used in the measurement with \mbox{amplitude $>\SI{100}{\milli\volt}$} added an unacceptable force background for which parallelism measurements could not be performed during actual force measurements.

We now implement a purely optical detection method to measure and keep parallelism at all times during the measurements. To achieve this, the fibers of the three lower interferometers are glued in place to the lower plate cylinder using epoxy resin 
and polished together with the plate to exactly the same height. After polishing, the fiber ends are protected using photo resist, and the metallic coating of thickness $d_c$ is applied to the plate. As $d_c$ can be measured independently, the separation between the plates $a_i=a+\delta a+d_c$ can be determined at the position of each interferometer $i$ after removal of the photo resist. The tilt values $\delta a$ can then be reduced using a feedback circuit and the three piezo-electric stages, similar as demonstrated previously~\cite{Sedmik:2018kqt}.

In order to estimate the error due to a relative angular deviation $\alpha$ between the plates for distance-dependent interfacial forces, we apply the Derjaguin approximation~\cite{Derjaguin:1934} for small $\alpha$.

\end{paracol}
\nointerlineskip
\begin{align}
M(a,\a)\approx\inv{A}\!\int\limits_A\!{\rm d}x{\rm d}y\,M(a+h(x,y))\>,\ \text{with } h(x,y)=x \tan \alpha\>,\text{ and }M(a)=\{P(a),\,P'(a)\}\>,
\end{align}
\begin{paracol}{2}
\switchcolumn
\noindent and we assume the tilt axis to coincide with the $y$-axis. Since the Casimir pressure gradient has the strongest distance dependence ($P_C'(a)\propto a^{-5}$) of all considered interactions, we take it as proxy for the error budget. In \figref{fig:tilt}, we plot the angles corresponding to a signal $\delta M=|M(a,\a)-M(a)|$ equal to the desired pressure and pressure gradient sensitivities of $\delta P=1\,$nN/m$^2$ and $\delta P'=1\,$mN/m$^3$, respectively. Note that a tilt of $\alpha=\SI{1}{\micro\radian}$ corresponds to a deviation of $6\,$nm at the fiber positions at the circumference of the lower plate, which is conveniently detectable. We, thus, set our target to $\alpha<\SI{0.5}{\micro\radian}$, which renders tilt negligible at all separations.
\begin{figure}[!hb]
    \includegraphics[scale=0.92]{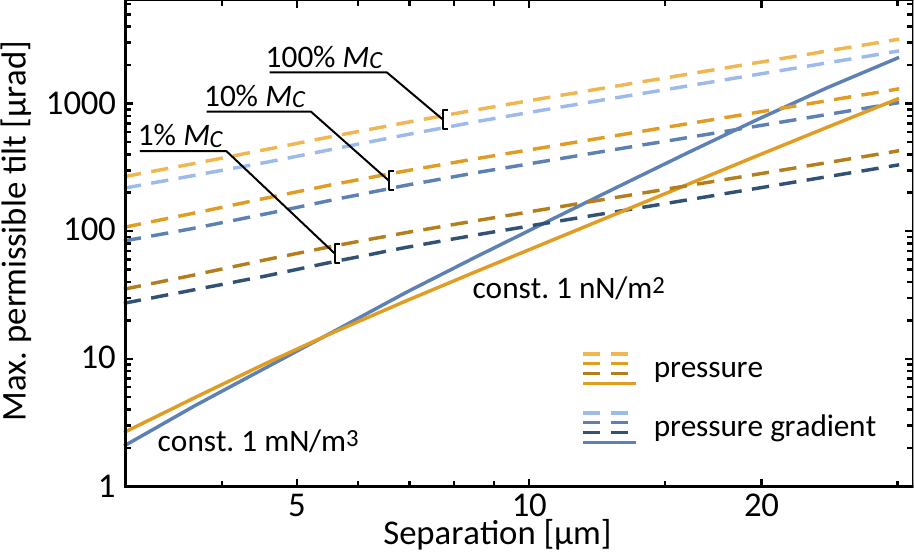}
    \caption{Maximum permissible tilt angle between the plates for an error of $\delta P=1\,$nN/m$^2$ for pressures and $\delta P'=1\,$mN/m$^3$ for pressure gradients (solid lines). The dashed lines indicate the angle at which the error $\delta M$ in the measured pressure or pressure gradient corresponds to the indicated fraction of the Casimir pressure or Casimir pressure gradient, respectively.\label{fig:tilt}}
\end{figure}

\subsection{Surface Flatness}
\label{sec:des:flat}
In the same way as tilt, corrugations of the surface change the effective distance $a_\text{eff}$ between the surfaces and, thus, introduce an error depending on the falloff of the measured interaction with the separation.
Again working in the Derjaguin approximation, we define
\begin{align}
    M(a)+\delta M_\text{corr}(a)=M(a_\text{eff})=\int\limits_{A}\!{\rm d}A\, M(a-h_1(x,y)-h_2(x,y))\label{eq:corrugation_error}\>,
\end{align}
for an arbitrary interaction $M(a)$ with measurement error $\delta M_\text{corr}$ due to corrugations with height profiles $h_1(x,y)$ and $h_2(x,y)$ on the two opposing surfaces. 
In order to obtain a realistic worst-case estimation for $\delta M_\text{corr}$, we consider two different forms of corrugations
\begin{enumerate}
    \item Stochastic short-scale roughness with RMS amplitude $h_s\ll a$. In this limit, the actual height distribution function $\rho_h$ can be used to statistically evaluate \mbox{\eqnref{eq:corrugation_error}}. While, for an actual experiment, the statistics of measured surface profiles have to be evaluated, we choose here a normal distribution around zero $\rho_{h,i}(\eta,h_s)=\sqrt{2\pi h_s^2}^{-1}\exp\left(-\eta^2/2h_s^2\right)$. The error then results from 
    \begin{align}
        \delta M_{\text{corr,s}}(a)&=\inv{N}\int\limits_{-h_\text{max}}^{h_\text{max}}\!{\rm d}\eta_1\, \rho(\eta_1,h_s)\int\limits_{-h_\text{max}}^{h_\text{max}}\!{\rm d}\eta_2\, \rho(\eta_2,h_s)M(a-\eta_1-\eta_2)\>,\\
        \quad\text{with } N&=\int\limits_{-h_\text{max}}^{h_\text{max}}\!{\rm d}\eta_1\, \rho(\eta_1,h_s)\int\limits_{-h_\text{max}}^{h_\text{max}}\!{\rm d}\eta_2\, \rho(\eta_2,h_s)\>,\nonumber
    \end{align}
    and $h_\text{max}$ is the peak roughness amplitude assumed to be equal on both plates.
    \item Large-scale deformations of the plates. For the lower plate of the recent experiment~\cite{Sedmik:2018kqt}, the surface was measured using an optical profilometer to have a roughly spherical deformation $h_1$ of depth 15\,nm (peak-peak) in negative direction (sag). It is technically, possible to obtain less than 10\,nm sag, and the actual deformations can be measured with $<$$1\,$nm accuracy. For the upper plate, we estimate the bend-through under gravity for a simply supported circular shell using the expression~\cite{Timoshenko:1959}
    \begin{align}
        h_2(r)=\frac{m g}{r_p^2\pi}\frac{r_p^2-r^2}{64 D}\left(\frac{5+\nu}{1+\nu}r_p^2-r^2\right)\>,\quad \text{with }D=\frac{E t^3}{12(1-\nu^2)}\,.\label{eq:plate_bending}
    \end{align}
    Here, $m\approx26\times 10^{-6}\,$kg is the mass of the upper plate, $g=9.81\,{\rm m}/{\rm s}^2$, \mbox{$r_p=5.742\,$mm} is the plate radius, $t=\SI{100}{\micro\metre}$ is the plate thickness, and we use the values \mbox{$E=170\,$GPa} and $\nu=0.07$ for the Young's modulus and Poisson ratio, respectively, of silicon along the $\langle 110\rangle$ axis~\cite{Hopcroft:2010}. \eqnref{eq:plate_bending} yields $z_{p,\text{max}}=13\,$nm maximum bend-through at the center ($r=0$). The error $\delta M_\text{corr}(a)$ is then computed using \eqnref{eq:corrugation_error}.
\end{enumerate}

Assuming typical values $h_s=1\,$nm and $h_\text{max}=10\,$nm, we obtain the error $\delta M_\text{corr,s}(a)$ shown in \figref{fig:corrugation_errors}, which is negligible at all separations. For the large-scale deformations of the plate, the influence is much more significant. In the previous experiment, the deformations of the upper and lower plates were both almost spherical with maximum deviations of $-13\,$nm and $-15\,$nm, respectively, at the center. These deformations, hence, almost cancel, yielding only a small relative change of separation. Despite this cancellation, the influence on the measured pressure is significant at small separations. For this reason, detailed and accurate measurements of the deformations $h_1(x,y)$ and $h_2(x,y)$ of both surfaces, as well as a detailed simulation of the gravitational sag of the upper plate, have to be performed. These measurements have to be taken into account during data analysis but represent a systematic error that can be subtracted. In order to estimate the residual error, we compute $\delta M_\text{corr}(a)$ for the conservative estimates $h_1=h_2=2\,$nm of purely spherical deformations.
\end{paracol}
\nointerlineskip
\begin{figure}[H]
\widefigure
    \centering
    \mbox{}\includegraphics[scale=0.88]{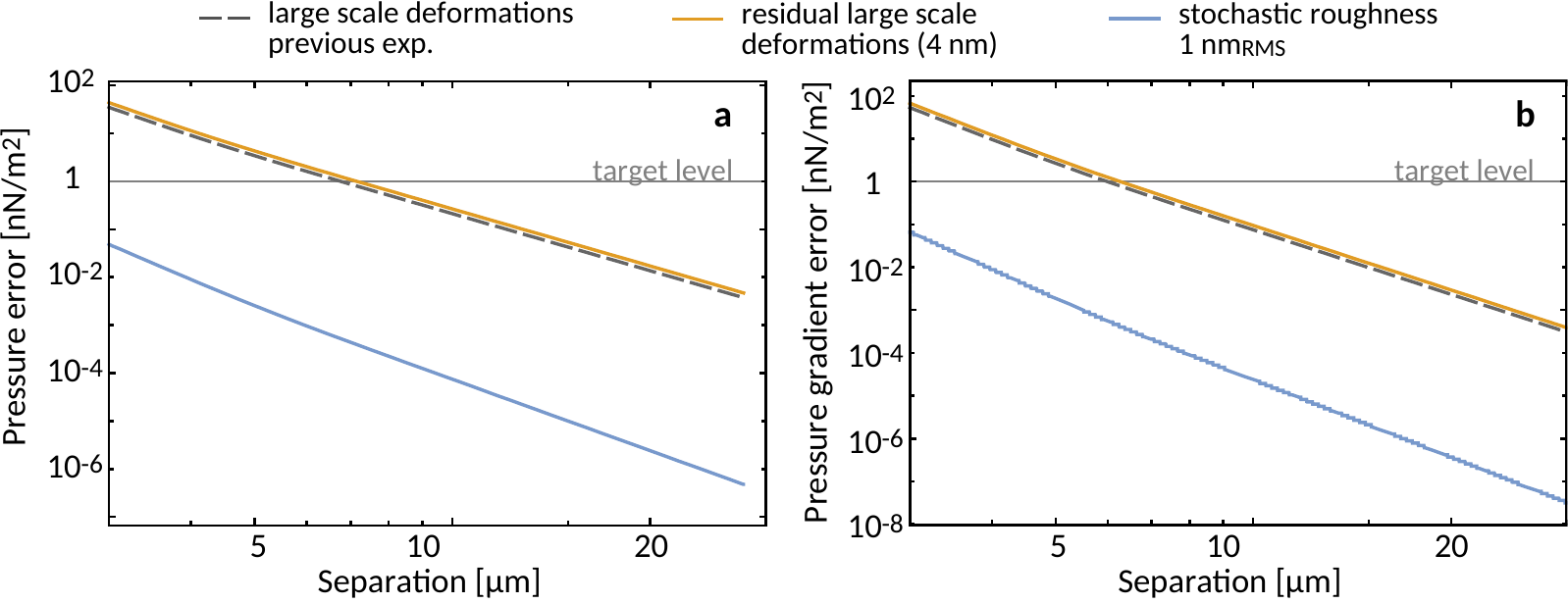}
    \caption{Errors $\delta P_\text{corr}$ of the pressure (\textbf{a}) and $\delta P'_\text{corr}$ of the pressure gradient (\textbf{b}) due to residual large scale deformations of the plates. For reference, the targeted sensitivity level is given by the gray line.}
    \label{fig:corrugation_errors}
\end{figure}
\begin{paracol}{2}
\switchcolumn

Finally, we need to address the influence of curvature (surface deformations) of the plates on the evaluation of prospective limits on fifth forces. Taking into account the measured and calculated deformations of the order \SI{10}{\nano\metre}, we obtain a radius of curvature on either plate of the order of $R_c\sim1\,$km. Concerning DM limits on Yukawa-like interactions, it has been shown in Reference~\cite{Decca:2009fg} that the \emph{proximity force approximation} (PFA) constitutes a valid approximation for such interactions and experimental setups with sphere-plate and plate-plate configurations. More recently, in Reference \cite{Elder:2019yyp}, the sphere-plate experimental configurations has been investigated with respect to symmetron DE for a sphere of radius $R$. It was found that, for $\mu\gtrsim10/R\sim10^{-9}\,$eV (in natural units), the PFA still provides a suitable tool for investigation. For the smallest value $\mu=10^{-2}\,$eV considered in \mbox{\secref{subsec:darkenergy}}, this requirement reads $R_c\gtrsim0.1\,$mm, which is clearly fulfilled. Consequently, for the expected sensitivity of \cannex{}, we trust the PFA to be a valid approximation to evaluate the influence of a possible curvature of the plates on the prospective limits for a large class of DM and DE candidates.

\subsection{Vibrations and Seismic Isolation System}
\label{sec:des:vibr}
Vibrations causing a fluctuation $\delta a$ in the plate separation and $\delta z_s$ in the sensor extension influence the achievable sensitivity in several ways. Firstly, the non-linearity of the measured pressure and pressure gradient with respect to the object separation causes a net offset in the measured signal. Secondly, the RMS noise linearly enters the measurements within the respective bandwidths. Thirdly, in measurements of the pressure, $\delta z_s$ is directly interpreted as a force on the sensor, while, for measurements of the pressure gradient, $\delta a$ leads to phase noise that enters the frequency measurement. The latter two effects reduce the signal-to-noise ratio (SNR) and the performance of various control systems (parallelism, distance, electrostatic compensation, calibrations), thereby increasing the systematic and stochastic error in non-trivial ways. The various errors are discussed in detail in \mbox{\appref{app:vibr_errors}}.

\begin{figure}[!ht]
    \mbox{}\includegraphics[scale=.9]{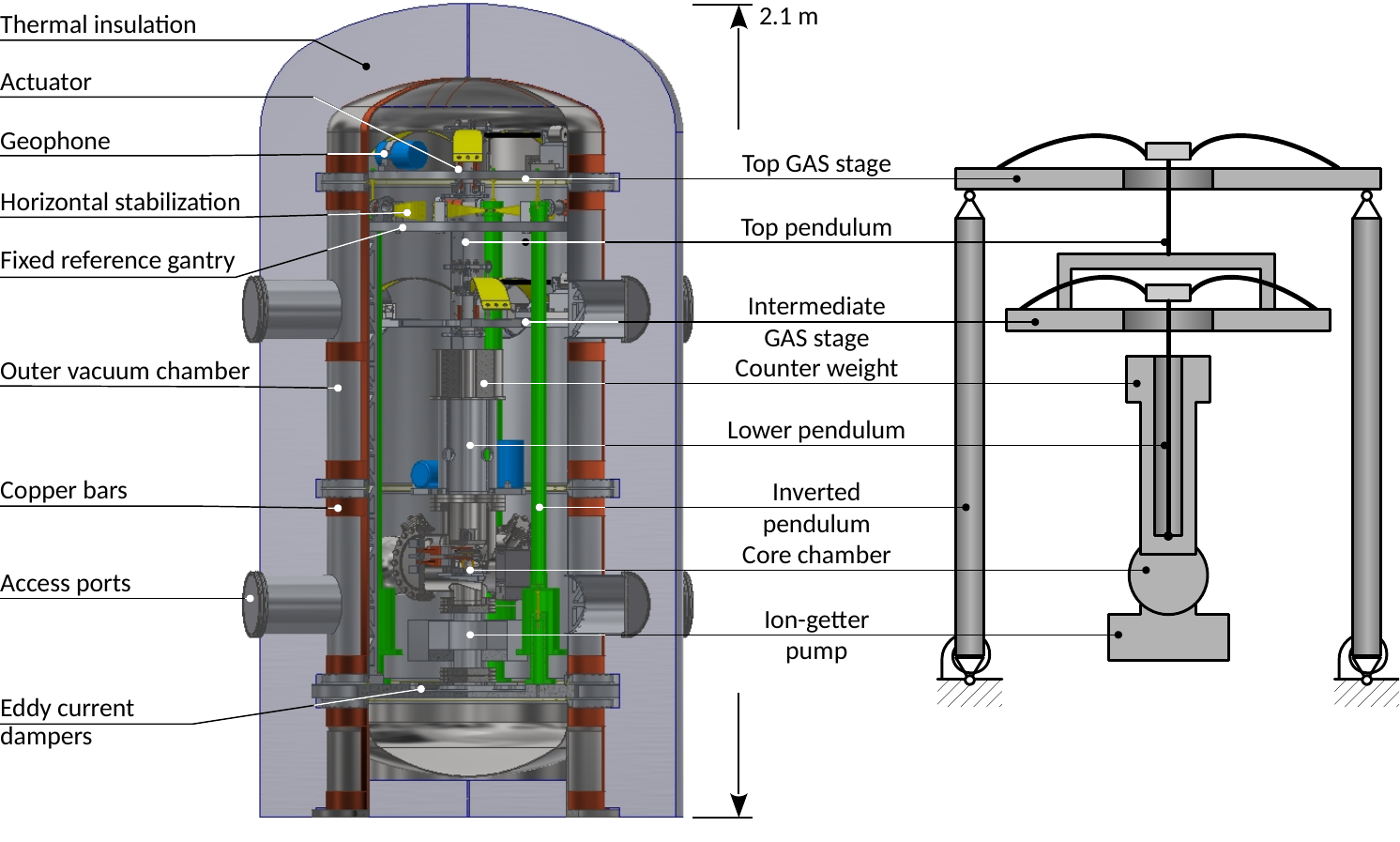}\mbox{}\vspace{-2pt}
    \caption{Overview of the \cannex{} setup. Left: Two-staged working version of the seismic attenuation and thermal insulation system. Right: Simplified schematic.\label{fig:sas}\vspace*{-10pt}\mbox{}}
\end{figure}
In order to reduce vibrations, \cannex{} uses technology originally developed for gravitational wave detectors~\cite{Braccini:2005hq,Stochino:2009zz}. As the performance of the previous one-staged active-passive seismic attenuation system (SAS) turned out insufficient~\cite{Sedmik:2018kqt}, we enhanced the design, as shown in \figref{fig:sas}. A very similar two-staged system was used for Advanced Virgo~\cite{Beker:2014zba}. For horizontal damping, we use an inverted pendulum~\cite{Takamori:2007zz} in combination with a double pendulum. Vertical attenuation is achieved by two sequential geometric anti-spring (GAS) filters~\cite{Cella:2004bn} with active feedback on the first stage~\cite{Mantovani:2005} similar as used previously~\cite{Sedmik:2018kqt}. We use \emph{magic wands}~\cite{Stochino:2007zz} and counter weights~\cite{Takamori:2007zz} to improve the attenuation at `high' frequencies around $\omega_0$. This latter measure also allows us to set the resulting zeros of transfer functions such that they reduce the main resonances of the second stage. 
\end{paracol}
\vspace{-7pt}
\begin{figure}[H]
\widefigure
    \centering
    \mbox{}\includegraphics[scale=.90]{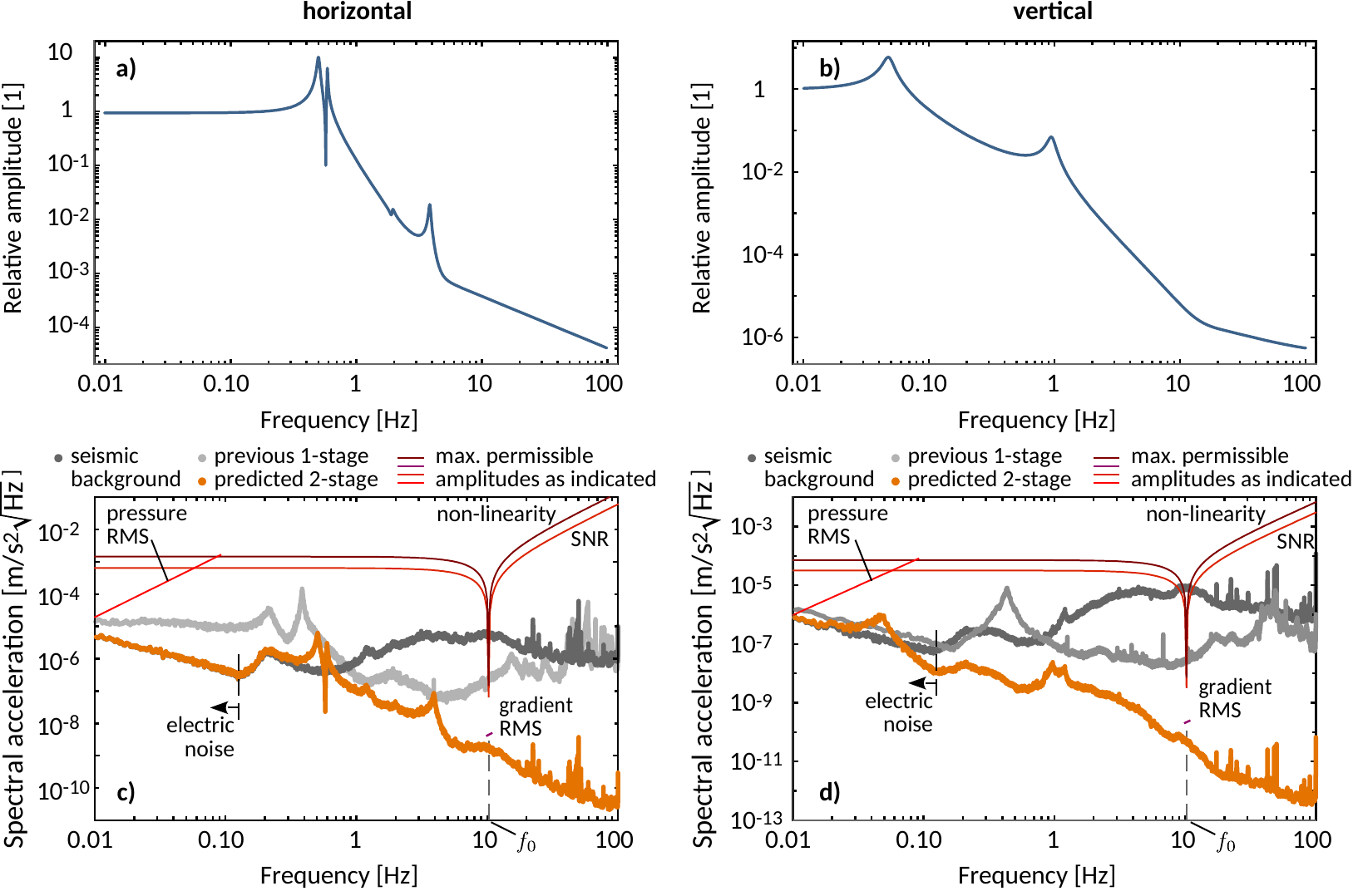}
    \caption{Preliminary predicted performance of the improved \cannex{} SAS. Left graphs: horizontal vibrations; right graphs: vertical vibrations. (\textbf{a}),(\textbf{b}): Computed spectral attenuation factor of the improved SAS. (\textbf{c}),(\textbf{d}): Measured spectral seismic acceleration background and predicted level for the core chamber. Maximum admissible amplitudes computed from requirements on the RMS noise, non-linearities, and the signal-to noise ratio (SNR) influencing measurements in a non-trivial way (discussed in the appendix) are given as red lines of different hue. Below $\sim\SI{0.1}{\hertz}$, electronic $1/f$ noise dominates the signal of our seismic measurements. The actual seismic level, thus, lies below the curves at these frequencies.\label{fig:seismic_performance}}
\end{figure}
\begin{paracol}{2}
\switchcolumn
The low-frequency behavior of this two-staged SAS can be modeled by means of lumped parameter models, as is described in the literature~\cite{Beker:2014zba}. In \mbox{\figref{fig:seismic_performance}a,b}, we show preliminary plots of the resulting attenuation in horizontal and vertical direction, respectively. As for a similar system, crosstalk between vertical and horizontal degrees of freedom has been shown to be insignificant~\cite{Beker:2014zba}, and 
we neglect such effects also here. In horizontal direction, visible resonances of the inverted, upper, lower pendulum, and combined pendula stages can be identified at frequencies \SI{0.6}{\hertz}, \SI{0.48}{\hertz}, \SI{4.0}{\hertz}, and \SI{1.9}{\hertz}, respectively. The zero of the inverted pendulum is set near $f_0$. In vertical direction, the visible resonances at \SI{0.05}{\hertz} and \SI{1}{\hertz} are again the result of a complex interplay between the two filters and the wands, while the zero of the second wand is clearly seen near $f_0$. Frequencies have been chosen to minimize overlap between horizontal and vertical resonances. Note that the effect of active feedback has not been included here, and optimization is still in progress. However, as can be seen in \figref{fig:seismic_performance}c,d, the predicted attenuation suffices to fulfill the requirements discussed above. At the resonance, an attenuation of $-71$ and $-106\,$dB in amplitude is achieved in horizontal and vertical direction, respectively. The predicted distances to the requirement at the resonance of 8 and 11\,dB, respectively, leave us confident that the requirements can also be met in the presence of unforeseen environmental or technical problems.

\subsection{Thermal Errors and Control}
\label{sec:des:thermal}
The previous version of \cannex{} was equipped with active thermal controls in the core, as well as on the surrounding vacuum chamber. However, after a flooding event, the thermal controllers were damaged, and heating cycles of the building caused variations of more than $1\,^\circ$C over the course of a day. For the redesign, we improved the setup in two~ways.

Firstly, the outer chamber is now equipped with a dense grid of copper bars with \SI{25}{\centi\metre} spacing, contacting the chamber wall via thermally conducting glue. On the grid, 50~independent and calibrated thermal controllers, each having its own sensor and Peltier-actuator regulate the local temperature. In addition, we increased the passive insulation (expanded polystyrene) thickness to \SI{20}{\centi\metre}. Using 3D numerical simulation, we computed the maximum differential temperature on the outer chamber wall to be \SI{7}{\milli\kelvin} for a change of \SI{1}{\celsius} in the ambient room temperature, while the average chamber temperature can be kept stable at \SI{1}{\milli\kelvin} level. This stability is clearly sufficient to avoid any detuning of the SAS. It remains to stabilize the core setup. The interferometric measurement of $a$ enables us to correct continuously for any error in separation or tilt due to thermal expansion. However, due to the large temperature coefficient $TCE=-64\times 10^{-6}$ of the Youngs modulus of Si~\cite{Hopcroft:2010}, temperature deviations $\delta T$ on the sensor need to be kept as small as possible. 
\begin{figure}[!hb]
    \includegraphics{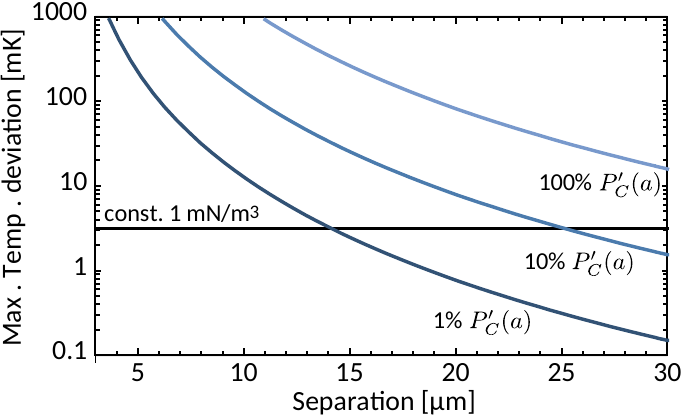}
    \caption{Maximum permissible thermal deviation of the sensor for a measurement of the Casimir pressure gradient at indicated precision or a constant pressure gradient sensitivity of $1\,$mN/m.\label{fig:thermal_dev}}
\end{figure}
If we consider as measurement error the difference $\delta P_T=P_m-P_{act}=P_{act}[(1+TCE\,\delta T)^{-1}-1]$ between measured $P_m$ and actual pressure $P_{act}$ between the plates, respectively, we obtain $\delta P_T<\SI{0.1}{\pico\newton/\centi\metre^2}$ even at shortest separation with electrostatic disturbances of \SI{10}{\milli\volt} and $\delta T=\SI{100}{\milli\kelvin}$. For gradient measurements, however, we have that $\delta k=k\,TCE\,\delta T=A\,\delta\left(\partial P_m/\partial a\right)$. In \figref{fig:thermal_dev}, we indicate the max. temperature deviation on the sensor for a $\delta P_T$ equaling 1\%, 10\%, and 100\% of the Casimir pressure at different separations, as well as constant error of $\delta P'(a)=1\,$mN/m$^3$. 
We, therefore, set our target for stabilization of the sensor temperature to $\delta T<0.1\,$mK.
In order to achieve this, both plates (as well as several parts of the core) have their own thermal controls~\cite{Klimchitskaya:2019nzu}. The lower plate has an integrated platinum sensor and a Peltier element attached to its lower end, allowing us to increase or decrease the temperature of the lower plate by $\pm\SI{10}{\degreeCelsius}$. In order to reduce thermal radiation to/from the plate from/to surrounding parts, not only the top surface but also its sides are coated with an optically thick layer of gold. Mechanical contact to the surrounding Macor holder is made on the side via three spring-loaded sapphire balls\footnote{This form of connection is similar to a jewel bearing. Further contact is made via the three glass fibers having low thermal conductance $\sim$$1.4\,$W/m\,K~\cite{Haynes:2014} and three thin wires establishing electrical contact that, however, are thermalized via the Peltier element.}. Excess heat is removed/delivered to the lower end of the Peltier element via a heat pipe leading out of the core (see \figref{fig:core_schematic}).

For the upper plate, no mechanical contact to thermal controls is possible. We, therefore, embed the frame of the sensor in an aluminium holder stabilized by a Peltier element and Pt resistors. To control the central part and the springs directly, we added a thermal copper shroud with Peltier actuator that is sputter-coated with graphite on the inside. The temperature of the sensor is measured by a thermopile sensor with 295 V/W and a noise floor of 18\,nV/$\sqrt{\rm Hz}$, which theoretically enables us to resolve 0.03\,mK variations on the~sensor.

The thermal noise of the mechanical resonator is strongly suppressed by the large $Q$-factor. At room temperature, we have~\cite{Saulson:1990} $|\delta P_B|A=\sqrt{4 k_B T m_\text{eff}\omega_0/Q}=0.07\,{\rm pN}/\sqrt{\rm Hz}$. Integrated over the bandwidth $f_{bw}$, this yields $3.9\,$fN for the minimally detectable force (or $38.6\,$pN/m$^2$ for the pressure), which is negligible compared to other sources of error.

\subsection{Electrostatic Patch Effects}
\label{sec:des:patches}
Electrostatic patches are local variations of the Fermi potential (or surface charge) exhibited by surfaces that have widely been investigated both theoretically~\mbox{\cite{Speake:2003zz,Kim:2009mr,Decca:2009,Behunin:2011gj,Behunin:2012zr}} and experimentally~\cite{Rossi:1992,Robertson:2006,Schafer:2016,Garrett:2020acf}, not only in the context of Casimir experiments. In particular, for the frequently used plate versus sphere geometry, patch effects have been shown to represent a limitation to the achievable accuracy~\cite{Behunin:2013qba}. Ref.~\cite{Behunin:2011gj} found that, for the limit $\lambda_p/a\to 0$, for patches of mean size $\lambda_p$, the patch pressure due to the RMS patch potential $V_\text{rms}$ obeys a simple scaling law $P_p\approx 0.9 \vare_0 V_\text{rms}^2\lambda_p^2/a^4$, while, in the opposite limit $\lambda_p/a\to\infty$, $P_p\propto 1/a^2$ results. As typical values for gold surfaces are $\lambda_p<500\,$nm~\cite{Decca:2005yk}, AFM and micro-oscillator experiments mostly operate in the latter regime where the resulting error from patches may be a nuisance. For parallel plates at large separation, where the Casimir pressure traverses from the $a^{-4}$ dependence to $a^{-3}$ in the thermal regime and $\lambda_p\ll a$, the situation relaxes, as has been reasoned in the literature~\cite{Sedmik:2018kqt,Klimchitskaya:2019fsm,Sedmik:2020cfj}. From actual Kelvin probe scans over $8\times8\,\si{\micro\metre^2}$, we inferred $V_\text{rms}=\SI{0.64}{\milli\volt}$ for the plates used in \cannex{}. Nonetheless, there are indications that patch distributions at larger scale may exist~\cite{Robertson:2006,Antonini:2008bb}. For this reason, we now implement directed plasma-cleaning via an Ar ion gun in our setup. This established technique~\cite{Hite:2012,Xu:2018} reduces the surface potential due to hydrocarbons, water, and other contaminants that are known to cause patches of larger diameter~\cite{Rossi:1992}. Ref.~\cite{Liu:2019xgg} improved the surface treatment further by adding UV irradiation as pre-cleaning step to lower the residual variation of the surface potential over distance (which is characteristic for patch effects) to less than $1\,$mV. Both techniques can be incorporated into \cannex{} as it is possible to increase the plate separation to \SI{5}{\milli\metre}, which enables direct ion impact at grazing incidence from a source outside the actual core (see \figref{fig:core_schematic}). The availability of high-power UV-LEDs allows us to also incorporate directed UV irradiation into our setup. In order to prevent re-deposition of volatile elements over time~\cite{Rossi:1992}, high vacuum $<10^{-7}\,$mbar is required for which \cannex{} employs an ion-getter pump that was tested to achieve and maintain a residual pressure of $10^{-9}\,$mbar in the core chamber. Due to the sensitive translator stages, force sensor, and optics, bakeout is not an option, though. Regarding the actual influence of patches, it was estimated in Refs.~\cite{Klimchitskaya:2019fsm,Sedmik:2020cfj}, using a model valid for $\lambda_p/a\ll 1$ and also the full theory~\cite{Behunin:2011gj}, that patch effects with $V_\text{rms}=0.64\,$mV are only relevant in \cannex{} at the targeted sensitivity level at $a\lesssim \SI{8}{\micro\metre}$, where Casimir interactions are stronger by several orders of magnitude.
\subsection{Detection Errors}
\label{sec:des:det}
\cannex{} has two simultaneous detection modes (pressure and pressure gradient), which both rely on the calibration of the sensor's mass $m_\text{eff}$, free resonance frequency $\omega_0$, the feedback-based cancellation of residual DC potentials between the plates (described in the appendices of Ref.~\cite{Sedmik:2018kqt}), and the parameters of the interferometric detection systems.

In order to estimate the systematic (non-subtractable and non-reducable by averaging) errors $\delta P'_\text{det}(a)$, we consider \eqnref{eq:freq_shift}, from which we obtain for the measured total pressure gradient $P'_\text{tot}(a)$ between the plates
\begin{align}
    P'_\text{tot}(a)=P'_{ES}(a)+P'_{C}(a)+P'_{G}(a)+P'_{\text{hyp}}(a)=\frac{A}{m_{\text{eff}}}(\Delta\omega^2-2\omega_0\Delta\omega)\label{eq:exp_press_grad}\,.
\end{align}

The gravitational and hypothetical interaction gradients $P'_G(a)$ and $P'_{\text{hyp}}(a)$, respectively, are negligible\footnote{Here, we preclude that any hypothetical force will be much smaller than the Casimir force, as the latter has been measured at the percent level without significant disturbance. However, we may not a priori exclude the possibility that a hypothetical additional force may be responsible for the discrepancy between data and predictions by Lifshitz theory with the Drude model discussed in \secref{subsec:casimir}.} with respect to the Casimir pressure gradient. Consequently, we extract $P'_C(a)=P'_\text{tot}(a)-P'_{ES}(a)$ by subtracting the known electrostatic gradient $P'_{ES}(a)=-\vare A [V_{ex}^2+V_{AC}^2+2 (V_{DC}+V_0)^2]/(2a^3)$. $V_{AC}$ is the excitation voltage amplitude of the electrostatic compensation circuit~\cite{Sedmik:2018kqt}, which effectively drives the applied $V_{DC}\to -V_0$ to cancel the global surface potential $V_0$, and $V_{ex}$ is the excitation amplitude of the PLL circuit. The error $\delta P'_\text{det}(a)$ is obtained by propagating the uncertainties of all parameters appearing in \eqnref{eq:exp_press_grad}. 

For measurements of the pressure, we have the expression
\begin{align}
    P_\text{tot}(a)=P_{ES}(a)+P_{C}(a)+P_{G}(a)+P_{\text{hyp}}(a)=\inv{A}\Delta z_s\omega_0^2 m\>,\label{eq:exp_press}
\end{align}
from which we obtain the Casimir pressure $P_C(a)$ under the same assumptions as described for the pressure gradient above, with the exception that gravity has to be considered here (see \secref{sec:des:total}). To calibrate the mass, we move to large separation $a_\text{cal}\approx\SI{100}{\micro\metre}$ (precisely determined by the interferometric distance measurement), set $V_{AC}\to 0$, and then record the frequency change for a range of applied $V_{DC}$ values~\cite{Sedmik:2018kqt}. Since, at such large separation, and especially for larger $V_{DC}$, all but the electrostatic interaction become negligible, we can determine $m$ by considering the measured electrostatic response in \eqnref{eq:exp_press_grad} only. The sub-percent contributions of other forces only result in an offset that can be fitted. Under the same assumptions, the free resonance frequency $f_0=\omega_0/(2\pi)$ can be measured using the PLL at $a=a_\text{cal}$ with activated $V_{AC}$ feedback and under consideration of the theoretical Casimir pressure gradient.

 It remains to compute the statistical errors coming directly from the interferometric detection systems. With respect to pressure gradients, the most critical parameter is the frequency resolution $\delta f=\SI{1}{\micro\hertz}$ of the PLL. For integration times $\tau\sim 600\,$s, the Allen deviation from the oscillator stability ($\delta f_\text{osc}/f_\text{osc}=2\times 10^{-10}$ for $\tau=0.1\,$s, according to manufacturer data) is at a comparable level for which there still remains some potential for optimization. We consider that we have three independent interferometers (and PLLs) determining $\Delta \omega$, and statistical averaging over $N=100$ independent measurements in the presence of noise. For DC measurements of the sensor extension $\Delta z_s$, the noise of the detector is critical. Considering a detector signal $V_\text{det.}=A+B\cos 4\pi z_s/\lambda$ for constants $A\approx B\sim 1\,$V of the single-fiber Fabry P\'erot interferometric measurement with cavity size $z_s\approx \SI{100}{\micro\metre}$, we obtain for the maximum permissible RMS detector noise voltage $\delta V_\text{det}=4\pi B\,\delta z_s/\lambda$. By demanding $\delta z_s<P_\text{lim}A/k$ for $P_\text{lim}=1\,$nN/m$^2$, we obtain $\delta V_\text{det}<\SI{6}{\micro\volt}\text{rms}$, which can be achieved using small bandwidth detection and long integration times. Another important parameter is the laser wavelength stability. Again, we consider the interferometer signal from which we obtain the requirement $\delta\lambda<P_\text{lim} \lambda A/(k z_s)$, which translates to $\delta\lambda<\SI{12}{\femto\metre}$. For commercially available frequency-stabilized C-band lasers at $\lambda\sim 1550\,$nm, linewidths $\delta f_\text{LW}$ around \SI{10}{\kilo\hertz} are specified, which results in $\delta\lambda=(\lambda^2/c)\delta f_\text{LW}=\SI{0.13}{\femto\metre}$. For frequency shift measurements, $\delta\lambda$ acts as a phase noise (see \appref{app:vibr_errors}), for which we find $\delta\lambda<\lambda^2QA/(4\pi a m_\text{eff}\omega_0^2)P'_\text{lim}=\SI{0.16}{\nano\metre}\times\SI{10}{\micro\metre}/a$ if we impose $P'_\text{lim}=1\,$mN/m$^3$. This is significantly more than exhibited by standard telecom laser sources, which have a line width of $\SI{10}{\mega\hertz}$ (corresponding to $\delta\lambda=\SI{80}{\femto\metre}$) and stability in the pm range. We remark that the error of frequency detection is predominantly given by the phase error and frequency resolution of the PLL, while the laser bandwidth only contributes at the nHz level to the measured $\Delta \omega$. Finally, voltage amplitudes are generally read back and evaluated by a lock-in amplifier with long integration times, leading to uncertainties at sub-\si{\micro\volt} level. 
\end{paracol}
\nointerlineskip
\begin{specialtable}[ht] 
\widetable
\caption{Parameters and their uncertainties/errors used for the analysis of the detection error $\delta M_\text{det}$ (IF = Interferometer).\label{tab:measurement_params}}
\scalebox{0.87}[0.87]{
\begin{tabular}{ccccc}
\toprule
\textbf{Parameter} &\multicolumn{1}{c}{\textbf{Value}}&\multicolumn{1}{c}{\textbf{Error}}& \textbf{Unit} & \textbf{Main Error Source}\\
\midrule
Sensor mass $m_\text{eff}$ & 31.748 & 0.003 &mg&{Vibrations and phase noise during calibration}\\
Sensor area $A$& 1.0834 & 0.0005& cm$^2$&Measurement uncertainty\\
Sensor extension $\Delta z_s$ & \multicolumn{1}{c}{variable} & 0.1 &pm&Laser linewidth, electronic noise.\\
Sensor free resonance $f_0$ & 10.243 & 2.$4\times10^{-8}$&Hz& Lower IF linewidth, electronic noise.\\
Sensor quality factor $Q$ & \multicolumn{1}{c}{$10^4$} & \multicolumn{1}{c}{-} & 1 &Residual gas\\
\midrule
Upper IF wavelength $\lambda_{DC}$ & 1590.0 & 1.$2\times 10^{-7}$ &nm & Laser stability\\
Upper IF linewidth $\delta f_\text{LW,DC}$ & \multicolumn{1}{c}{15} & \multicolumn{1}{c}{-}&kHz& -\\
Upper IF detector noise (RMS) & \multicolumn{1}{c}{1} & \multicolumn{1}{c}{-}&\si{\micro\volt}& -\\
Lower IF wavelength $\lambda_{AC}$ & 1550.0 & 8.$1\times 10^{-5}\,$ & nm& Laser stability\\
Lower IF linewidth $\delta f_\text{LW,AC}$ & \multicolumn{1}{c}{10} & \multicolumn{1}{c}{-}&MHz& -\\
Lower IF detector noise (RMS) & \multicolumn{1}{c}{100} & \multicolumn{1}{c}{-}&\si{\micro\volt}& - \\
PLL frequency resolution $\Delta f_\text{min}$ & \multicolumn{1}{c}{1} & \multicolumn{1}{c}{-}&\si{\micro\hertz}& -\\
PLL phase noise (RMS) $\delta\varphi_\text{PLL}$ & \multicolumn{1}{c}{1} & \multicolumn{1}{c}{-} & \si{\micro\radian}& - \\
\midrule
Excitation voltage ampl. $V_\text{ex}$ & \multicolumn{1}{c}{$1\times a/\SI{10}{\micro\metre}$} & \multicolumn{1}{c}{$5\times 10^{-4}$} & mV& Measurement uncertainty\\
Voltage feedback ampl. $V_\text{AC}$ & \multicolumn{1}{c}{$2\times a/\SI{10}{\micro\metre}$} & \multicolumn{1}{c}{$5\times 10^{-4}$} & mV& Measurement uncertainty\\
Effective sensor vibr. ampl. (RMS, $\tau=600$\,s) $\delta z_c$ & 0.106 & \multicolumn{1}{c}{-} & pm& Measurement uncertainty\\
\bottomrule
\end{tabular}}
\end{specialtable}
\begin{paracol}{2}
\switchcolumn
 Values for parameter errors were determined during the recent experimental campaign. However, these errors can be lowered, if the improved SNR of the new setup is considered. From the improved SNR, previous measurements, and manufacturer data, we obtain the values and errors collected in \tabref{tab:measurement_params}, which we consistently propagate through \eqnsref{eq:exp_press} and \eqref{eq:exp_press_grad} to determine calibration and measurement errors. In \figref{fig:det_err_contrib}, we show the most significant contributions to $\delta M_\text{det}$. For pressures, this is the determination of the sensor extension which, in turn, receives contributions from vibrations, laser linewidth, and voltage noise of the detector and the data acquisition. For the pressure gradient, the most important error comes from the frequency measurement, which is given by the PLL's frequency resolution and phase noise, as well as vibrational phase noise. We note that the temperature-dependence of device errors has not been considered here, as it can always be reduced by providing a properly controlled environment. We also omit here errors for potential measurements at higher ambient pressures that have been proposed for the detection of DE chameleon interactions~\cite{Brax:2010xx}. While the resulting changes in the optical cavity sizes can be measured and corrected for by using an independent fixed-size Fabry P\'erot cavity, the lower Q-factor mainly increases Brownian noise.

\end{paracol}
\nointerlineskip
\begin{figure}[t]
\widefigure
    \centering
    \mbox{}\includegraphics[scale=0.86]{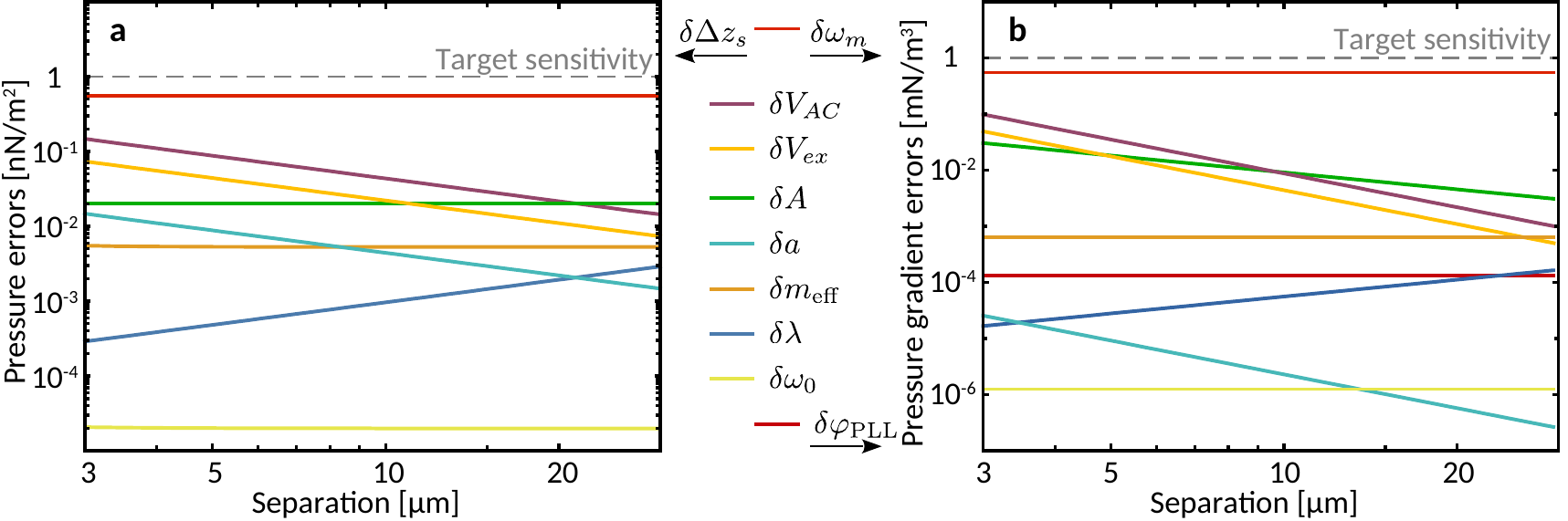}
    \caption{Contributions of the different parameter errors to the detection errors $\delta P_\text{det}$ (left) and $\delta P'_\text{det}$ (right). Additional contributions below the vertical scales exist. \label{fig:det_err_contrib}}
\end{figure}
\begin{paracol}{2}
\switchcolumn

\subsection{Total Realistic Error Estimate}
\label{sec:des:total}
For the total error budget, we consider all errors discussed in Sections \ref{sec:des:parallel}--\ref{sec:des:det}. We assume thermal $\delta M_\text{th}$, Brownian $\delta M_\text{B}$, seismic $\delta M_\text{seis,stat}$, and tilt errors $\delta M_\text{tilt}$ as statistically independent random contributions and residual electrostatic patches $\delta M_\text{patch}$, surface deformations $\delta M_\text{corr}$, non-linearity seismic $\delta M_\text{seis,sys}$ (see \appref{app:vibr_errors}), and detection uncertainty $\delta M_{det}$ as systematic errors that do not reduce with integration time but cannot be simply subtracted from the data due to unknown quantities. As the present analysis represents only prospective performance parameters and errors, we assume normal distributions throughout and state expected errors at the 68\% confidence level. The errors are shown together with the Casimir interaction (based on extrapolation of the dielectric function to zero frequency with the plasma model), the strength of the electrostatic excitation, and the gravitational interaction between the plates in \figref{fig:budget}.

\end{paracol}
\nointerlineskip
\begin{figure}[H]
\widefigure
    \centering
    \mbox{}\includegraphics[scale=0.86]{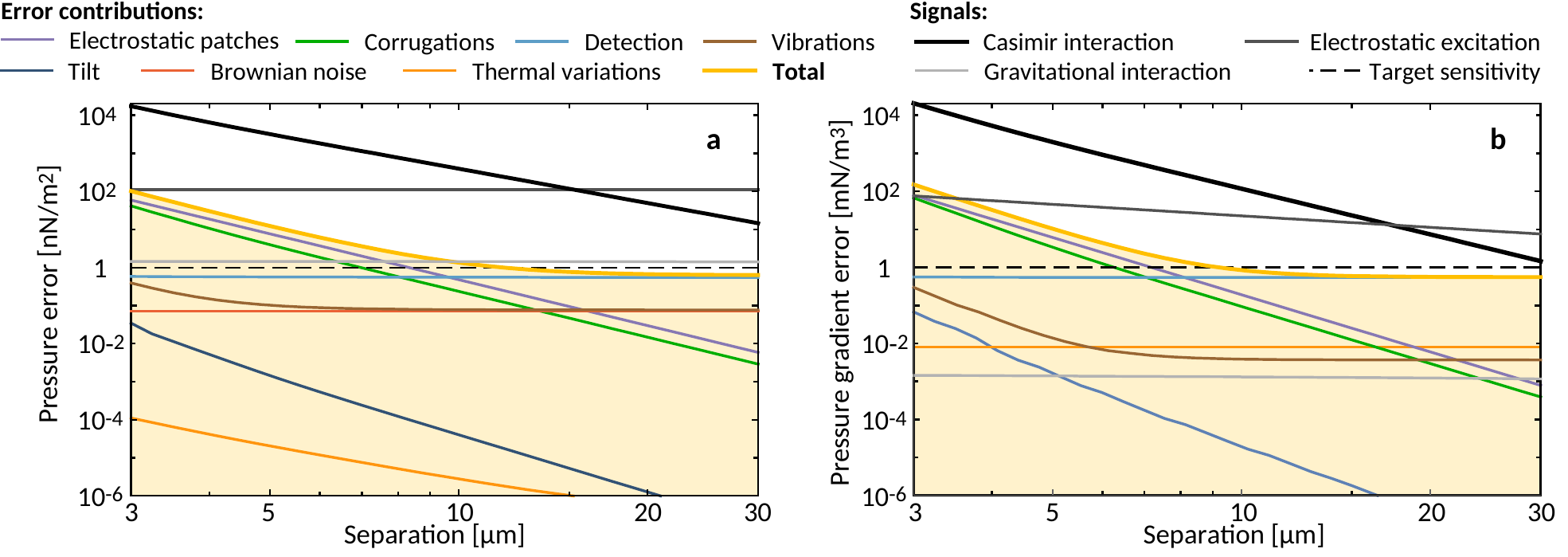}
    \caption{Error budget for the next-generation design of \cannex{} for measurements of the pressure (\textbf{a}) and pressure gradient (\textbf{b}).}
    \label{fig:budget}
\end{figure}
\begin{paracol}{2}
\switchcolumn

\textls[-15]{Measurements at or below the targeted sensitivity level clearly are possible at \mbox{$a>\SI{10}{\micro\metre}$}.} At smaller separations, the assumed large-scale corrugations of the plates and residual patch effects represent the major disturbances. We note that corrugations have to be evaluated carefully on the actually produced plates and sensors. Ideally, they can be taken into account during data evaluation, thereby strongly reducing their contribution to the error. The uncertainties considered here are, therefore, to be considered as pessimistic estimates. For electrostatic patches, the effect of combined UV irradiation and Ar ion cleaning still has to be evaluated. Such analysis has to be performed \emph{in situ}, as any exposure of the plates to air may alter the surface potentials. We, therefore, aim to perform Kelvin probe measurements on the lower plate by replacing the sensor by a custom AFM based on ferrule-top technology~\cite{Chavan:2010}. Again, the present estimation can be considered to be pessimistic with potential for reduction in the experiment. At all separations, the measurement sensitivity is presently limited by several factors (see \secref{sec:des:det}) to $0.55\,{\rm nN/m}^2$ and $0.53\,{\rm mN/m}^3$. This sensitivity, however, suffices to perform a direct measurement of the gravitational interaction between the two plates amounting to $1.43\,{\rm nN/m}^2$ at $a=\SI{15}{\micro\metre}$, but almost independent of separation. The gravity gradient in \figref{fig:budget}b remains beyond the bounds of the proposed experiment.

\section{Prospects}
\label{sec:prospects}
Prospects for \cannex{} have been presented in a number of articles~\cite{Almasi:2015zpa,Klimchitskaya:2019fsm,Klimchitskaya:2019nzu,Sedmik:2020cfj}. The basic assumption of sensitivities $1\,$nN/m$^2$ for pressures and $1\,$mN/m$^3$ for pressure gradients at separations around \SI{10}{\micro\metre} is reaffirmed for the new design. We, therefore, only briefly review existing prospective limits and discuss new prospective calculations for symmetron DE in \secref{subsec:darkenergy}.
%
%
\subsection{Casimir Interactions}
\label{subsec:casimir}
The comparably large pressure (gradient) sensitivity enabled by the use of the parallel plate geometry prospectively allows us to measure the Casimir effect with reasonable accuracy at separations larger than the thermal wavelength. This would enable an unambiguous characterization of the thermal Casimir effect in several ways.

The first opportunity regards the thermal Casimir force at large separation. In \mbox{\figref{fig:prospect_dp}}, we show computational results for the Casimir pressure and pressure gradient between parallel plates using the Drude, plasma, and non-local Drude approaches in units of the predicted total experimental error $\delta M_\text{tot}$. These results, thus, are indicative for the expected confidence level in the comparison between models. In these computations, we use tabulated optical data of Ref.~\cite{Palik:1998}, and the values $\omega_p=1.28\times 10^{16}\,{\rm s}^{-1}$, and $\gamma=7.14\times 10^{13}\,{\rm s}^{-1}$~\cite{Olmon:2012} for extrapolations of the optical data to zero frequency by means of the Drude and plasma approaches. For the non-local Drude model, we assume $\zeta^T=7$ and $\zeta^L=1$. For both, the pressure and its gradient, a clear distinction between the forces predicted by the Drude and plasma approaches is possible at all separations. We obtain $\max |P_{C,\text{Drude}}(a)-P_{C,\text{plasma}}(a)|/\delta P_\text{tot}=158$ at $a=\SI{7.8}{\micro\metre}$ and $\min |P_{C,\text{Drude}}(a)-P_{C,\text{plasma}}(a)|/\delta P_\text{tot}=11$ at $a=\SI{30}{\micro\metre}$ for the pressure in \figref{fig:prospect_dp}a. For the pressure gradient in \figref{fig:prospect_dp}b, the SNR is slightly worse with $\max |P'_{C,\text{Drude}}(a)-P'_{C,\text{plasma}}(a)|/\delta P'_\text{tot}=107$ at $a=\SI{6.8}{\micro\metre}$ and $\min |P'_{C,\text{Drude}}(a)-P'_{C,\text{plasma}}(a)|/\delta P'_\text{tot}=1.3$ at $a=\SI{30}{\micro\metre}$. Thus, an experimental exclusion of either model with high confidence ($>99$\%) seems possible for the full range of separations. With respect to a distinction between non-local Drude and plasma approaches, we have $\max |P_{C,\text{NLDrude}}(a)-P_{C,\text{plasma}}(a)|/\delta P_\text{tot}=40.7$ at $a=\SI{6.8}{\micro\metre}$ and $\max|P'_{C,\text{NLDrude}}(a)-P'_{C,\text{plasma}}(a)|/\delta P'_\text{tot}=32.7$ at $a=\SI{6.2}{\micro\metre}$. For both the pressure and pressure gradient, these values remain roughly constant for smaller separations but fall off quickly for larger separations. Hence, \cannex{} could provide experimental data that allows to clearly distinguish between proposed new non-local descriptions of the dielectric response and the plasma model.
\begin{figure}[H]
    \mbox{}\includegraphics[scale=0.8]{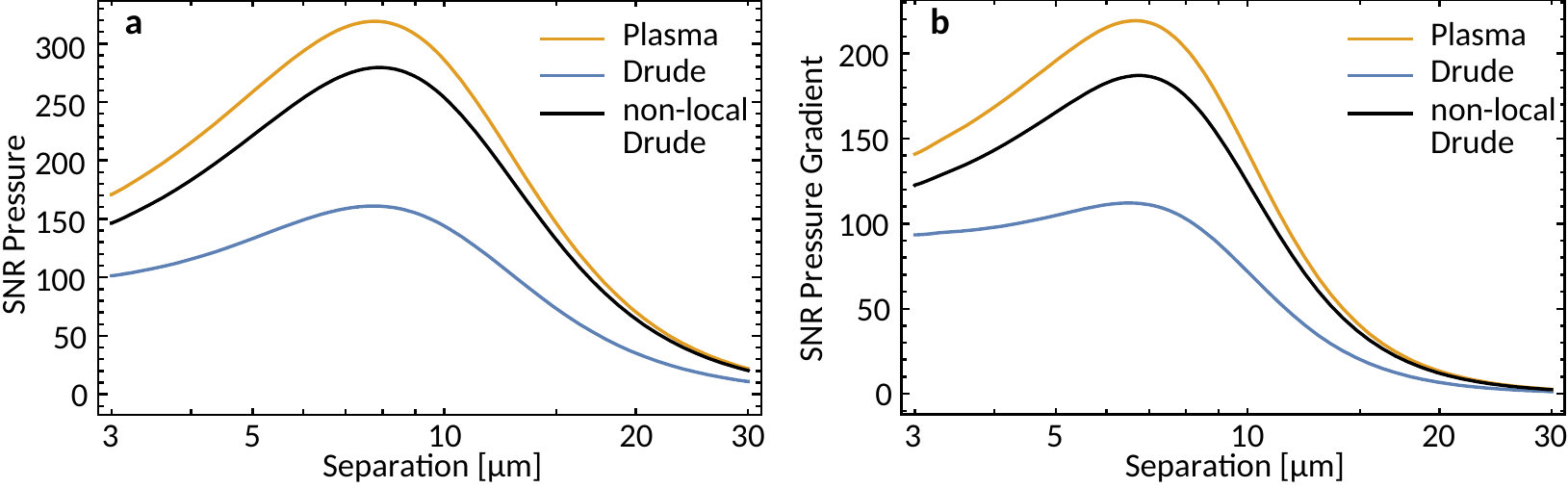}
    \caption{Casimir pressure (\textbf{a}) and pressure gradient (\textbf{b}) in units of the experimental error. Curves have been computed as described in \secref{sec:vacuum} for two gold semispaces using either the Drude model or plasma model to extrapolate tabulated optical data to zero frequency, or the non-local Drude model without consideration of optical data throughout.\label{fig:prospect_dp}}
\end{figure}

Recently, in Reference \cite{Klimchitskaya:2019nzu}, it was proposed to test the role of relaxation properties of conduction electrons in the Casimir pressure employing parallel plates, which are kept at different temperatures.
The total pressure for the plates consists in this case of three separate contributions. First, $P_\text{eq}(a,T)$ is the pressure related to thermal equilibrium for both plates at the same temperature $T$ (which is obtained as described in \secref{sec:vacuum} from \eqnref{eq:lifshitzT0} using the Matsubara formalism).
\begin{align}
\label{eq:Peq}
  P_\text{eq}(a,T) = - \frac{k_BT}{8\pi a^3}{\sum_{n=0}^\infty}'\!\int_{\zeta_n}^\infty\! {\rm d}y\,y^2\left\{\!\left(\frac{e^y}{r_{TM}^2(1,i\xi_n)} - 1\right)^{\!-1}\!+ \left(\frac{e^y}{r_{TE}^2(1,i\xi_n)} - 1\right)^{\!-1}\right\}.
\end{align}

Here, we use abbreviations for the complex frequencies 
\begin{align}
  \zeta_n = \frac{2a\xi_n}{c} = \frac{4\pi ak_BT n}{\hbar c}\>,
\end{align}
and $y=2a\sqrt{k_\perp^2 + \xi_n^2/c^2}$, where the magnitude of the transversal part of the wave-vector is denoted by $k_\perp$. Furthermore, $r_{TE}$ and $r_{TM}$ are defined in \eqnref{eq:Fresnel} with $\mu=1$. \eqnref{eq:Peq} assumes infinitely extended plates on either side of the vacuum gap, which constitutes a very good approximation for $P_\text{eq}(a,T)$ in the case of \cannex. The second contribution to the pressure is given by the expression
\end{paracol}
\nointerlineskip
\begin{align}
  &\Delta P_\text{neq}(a,T_1,T_2) = \frac{\hbar c}{64\pi^2a^4}\int_0^\infty \!{\rm d}u\,u^3\left(n(u,T_1) - n(u,T_2)\right)\sum_{\a} \nonumber\\
  &\times\Bigg\{\int_0^1\!{\rm d}t\,t\sqrt{1-t^2}\,\frac{|R_\alpha^{(2)}(u,t)|^2 - |R_\alpha^{(1)}(u,t)|^2}{|D_\a(u,t)|^2} \nonumber\\
  &- 2\int_1^\infty\!{\rm d}t\,t\sqrt{t^2 - 1}e^{-u\sqrt{t^2-1}}\,\frac{\text{Im}[R_\alpha^{(1)}(u,t)]\text{Re}[R_\alpha^{(2)}(u,t)] - \text{Re}[R_\alpha^{(1)}(u,t)]\text{Im}[R_\alpha^{(2)}(u,t)]}{|D_\a(u,t)|^2}\Bigg\},
\end{align}
\begin{paracol}{2}
\switchcolumn

\noindent where the sum denotes summation over the $TM$ and $TE$ modes indexed by $\alpha$ and
\begin{align}
  n(u,T) &= \left[\exp\left(\frac{\hbar c u}{2ak_BT}\right) - 1\right]^{-1} \>, \nonumber\\
  D_\a(u,t) &= 1 - R_\a^{(1)}(u,t)R_\a^{(2)}(u,t)e^{iu\sqrt{1 - t^2}}\>,
\end{align}
as well as
\begin{align}
  R^{(j)}_\alpha(u,t) = \frac{r_\alpha(1,\varepsilon_m) + r_\alpha(\varepsilon_m,\varepsilon_j)e^{2id_jk(\varepsilon_m)}}{1 + r_\alpha(1,\varepsilon_m)r_\alpha(\varepsilon_m,\varepsilon_j)e^{2id_jk(\varepsilon_m)}}\>.
\end{align}

The upper index on $R^{(j)}$ specifies the upper $(j=1)$ or lower $(j=2)$ plate, $\vare_m=\vare_m(\ri\xi_l)$ is the dielectric function of the gold coating (on both plates), and $\vare_j=\vare_j(\ri\xi_l)$ stands for the dielectric function of the substrates (Si for the upper plate and SiO$_2$ on the lower plate). Note, however, that the $\vare_j$ do not influence the results significantly, as the thicknesses $d_j$ of the metal coatings both are much larger than the electromagnetic penetration depth of gold. As $\Delta P_\text{neq}$ would vanish for equally thick plates, we choose values $d_1=\SI{200}{\nano\metre}$ and $d_2=\SI{1}{\micro\metre}$ that differ from the ones assumed in other calculations in this article.
Finally, the third contribution to the pressure stems from the Stefan-Boltzmann law
\begin{align}
  P_\text{SB} = \frac{2\sigma}{3c}(T_1^4 + T_2^4)\>,
\end{align}
describing the pressure on the outsides of the plates, which are in thermal exchange with an external heat bath at temperature $T_1$. As found in Reference \cite{Klimchitskaya:2019nzu}, already, 
a small temperature difference of only 10 K between the plates of \cannex{} could discriminate at high confidence between different theoretical predictions for the total pressure and its gradient, as well as between the contributions to them due to thermal nonequilibrium. Of particular interest is the case where the upper plate is kept at a fixed temperature $T_1$ equal to the environmental temperature, while the lower plate is heated to a slightly higher (or lower) temperature $T_2$. In this case, the total pressure on the upper plate $1$ is given by 
\begin{align}
  P_\text{tot}^{(1)}(a,T_1,T_2) = \frac{1}{2}\left(P_\text{eq}(a,T_1) + P_\text{eq}(a,T_2)\right) + \Delta P_\text{neq}(a,T_1,T_2) + \frac{2\sigma}{3c}(T_2^4 - T_1^4)\>.
\end{align}

For further details, we refer to Reference \cite{Klimchitskaya:2019nzu}.

 In \figref{fig:thermal_neq}, we show computational results for the three different contributions to $P_\text{tot}^{(1)}(a,T_1,T_2)$ for $T_1=300\,$K and $T_2=310\,$K. The equilibrium pressures and pressure gradients for different temperatures, as well as the thermal radiation pressure, clearly can be measured. We note that the difference in $P_\text{eq}$ between $T_1$ and $T_2$ amounts to $13\,\delta P(a)$ and $10\,\delta P(a)$ at $a=\SI{10}{\micro\metre}$ for the Drude and plasma approaches, respectively. For the pressure gradient, the differences in $P'_\text{eq}$ between $T_1$ and $T_2$ amount to $2.3\,\delta P'(a)$ and $4.7\,\delta P'(a)$ for the Drude and plasma approaches, respectively. The assumed thicknesses $d_U=\SI{200}{\nano\metre}$ on the upper and $d_L=\SI{1}{\micro\metre}$ on the lower plate result in contributions $\Delta P_\text{neq}$ and $\Delta P'_\text{neq}$ that only slightly exceed the predicted experimental error and are more than two orders of magnitude smaller than the total pressure and pressure gradient. By choosing a thinner layer on one side of the gap and larger temperature differences, however, these contributions would be enhanced~\cite{Ingold:2020rek} and could possibly be measured\footnote{Note that, due to time constraints during manuscript preparation, we show in \figref{fig:thermal_neq} only $\Delta P_\text{neq}$ and $\Delta P'_\text{neq}$ calculated using the Drude model, rather than the plasma model.}.
 \begin{figure}[!ht]
 \mbox{}\includegraphics[scale=0.86]{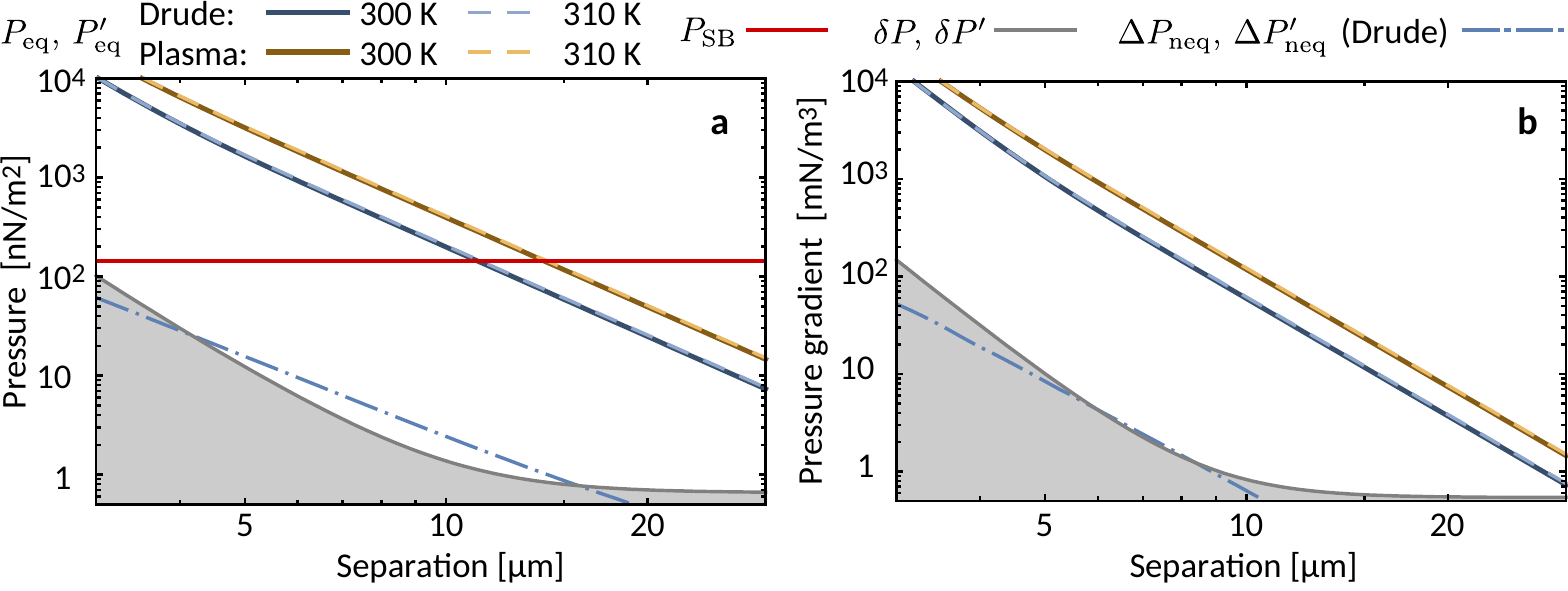}
 \caption{Comparison of the magnitude of the various contributions to the Casimir pressure (\textbf{a}) and its gradient (\textbf{b}) out of thermal equilibrium for temperatures $T_1=300\,$K and $T_2=310\,$K of the upper and lower plates, respectively. The experimental errors $\delta P$ and $\delta P'$ are given for reference.\label{fig:thermal_neq}}
\end{figure}

Based on the analysis in \secref{sec:des:total}, we may speculate that the same sensitivity predicted for gold surfaces could also be obtained for semi-conducting materials (with a thin gold layer on top). Therefore, the recent investigations regarding the role of free charge carriers on either side of the Mott-Anderson phase transition~\cite{Chen:2006zz} could be continued with higher experimental precision. Such measurement would clearly allow us to distinguish between the Drude or plasma prescriptions for the metallic phase. In addition, without any change in the setup, using the UV-LEDs that were added to reduce electrostatic potentials, we might be able to trigger the Mott-Anderson phase transition in a semiconductor with intrinsic charge carrier density slightly below the critical one~\cite{Chen:2007kqw}. This would give us the opportunity to investigate the observed change in the Casimir interaction in great detail. Eventually, any technically feasible material combination including magnetic materials, meta-materials, or graphene could be used, as long as surface potentials remain either constant such that they can be measured \emph{in situ} or can be reduced by using the UV/Ar ion cleaning method.

%
\subsection{Dark Matter}
\label{subsec:darkmatter}
As mentioned in \secref{sec:ODSIDM}, hypothetical new particles typically induce new interactions. In the non-relativistic limit, those interactions can be described in terms of effective potentials, which, in turn, can be probed via the experimentally observable effects they induce. 
In order to relate the effective fermion-fermion potentials \eqnsref{eq:scalar-scalar}, \eqref{eq:pseudoscalar-scalar}, \eqref{eq:vector-vector}, and \eqref{eq:axialvector-vector} to the experimentally measured pressure in the \cannex{} setup, summations over all interacting fermions of the parallel plates have to be performed. Each plate consists of several layers with individual finite thicknesses and varying fermion densities as outlined in \secref{sec:design}. This is done in Reference~\cite{Pitschmann:2020ejb}, where the induced pressure for two parallel plates with thicknesses $D_1$ and $D_2$ has been derived from the respective effective potentials. In this analysis, the plates have been considered as having infinite extension in both transversal directions, which constitutes a very good approximation for \cannex{}. For the effective potentials \eqnsref{eq:scalar-scalar} and \eqref{eq:vector-vector}, the pressures
\begin{align}
\label{eq:pr}
  \mathcal P_{SS}(a; \rho_{1}, \rho_{2}, D_1, D_2) &= -\frac{g_S^2\hbar^3}{2m_\phi^2c}\,\rho_1\rho_2\,e^{-a/\lambda}\,\mathfrak D(D_1,D_2)\>, \nonumber\\
  \mathcal P_{VV}(a; \rho_{1}, \rho_{2}, D_1, D_2) &= - \frac{g_V^2}{g_S^2}\,\mathcal P_{SS}(a; \rho_{1}, \rho_{2}, D_1, D_2)\>
\end{align}
have been obtained, respectively. Here, $\rho_1$ and $\rho_2$ have dimensions of an inverse volume and denote the average number of interacting fermions per unit volume in the respective plates, and the function $\mathfrak D(D_1,D_2)$ is given by~\cite{Pitschmann:2020ejb}
\begin{align}
  \mathfrak D(D_1,D_2) = \left(1 - e^{-D_1/\lambda} - e^{-D_2/\lambda} +  e^{-(D_1 + D_2)/\lambda}\right).
\end{align}

The \cannex{} setup consists of several layers and the total pressure is given by the sum of the individual pressures due to the pairing of each of the upper plate layers with each of the layers in the lower plate. In Reference~\cite{Pitschmann:2020ejb}, the total pressure for a more general setup consisting of $N_U$ homogeneous layers with fermion densities $\rho_{U,i}$ and thicknesses $D_{U,i}$ in the upper plate and $N_L$ homogeneous layers having densities $\rho_{L,j}$ and thicknesses $D_{L,j}$ in the lower plate has been obtained as
\end{paracol}
\nointerlineskip
\begin{align}\label{eq:TP1}
  \mathcal P_{SS}^\text{total}(a) &=  \sum_{i=1}^{N_U}\sum_{j=1}^{N_L} \mathcal P_{ss}\bigg(a + \sum_{i'=0}^{i-1} D_{U,i'} + \sum_{j'=0}^{j-1} D_{L,j'}; \rho_{U,i}, \rho_{L,j}, D_{U,i}, D_{L,j}\bigg) \nonumber\\
  &= -\frac{g_S^2\hbar^3}{2m_\phi^2c}\sum_{i=1}^{N_U}\sum_{j=1}^{N_L} \rho_{U,i} \rho_{L,j}\,\exp\bigg\{-\lambda^{-1}\bigg(a + \sum_{i'=0}^{i-1} D_{U,i'} + \sum_{j'=0}^{j-1} D_{L,j'}\bigg)\bigg\}\,\mathfrak D(D_{U,i}, D_{L,j}) \nonumber\\
  &= -\frac{g_S^2 \hbar^3}{2 m_\phi^2 c}\Phi(\lambda)\>,\nonumber\\
  \mathcal P_{VV}^\text{total}(a) &= - \frac{g_V^2}{g_S^2}\,\mathcal P_{SS}^\text{total}(a)\>, 
\end{align}
\begin{paracol}{2}
\switchcolumn

\noindent where $D_{U,0} = D_{L,0} = 0$. It is important to note that, in order to relate the effective fermion-fermion potentials, 
\eqnsref{eq:scalar-scalar}, (\ref{eq:pseudoscalar-scalar}), (\ref{eq:vector-vector}), and (\ref{eq:axialvector-vector}) to the corresponding induced pressures additional model assumptions have to be made. Specifically, those assumptions enter via the densities $\rho_{U,i}$ and $\rho_{L,j}$, which denote the average number of \emph{interacting} fermions per unit volume. For a specific model, like, e.g., the axion, its coupling to matter and, as such, $\rho_{U,i}$ and $\rho_{L,j}$ are well defined. While it couples via quarks to nucleons, there is to leading order no coupling to the electrons of the experimental setup's atoms. However, to compare bounds on axions with those for generic ALPs, one has to provide additional assumptions on the ALP matter coupling. Assuming a universal coupling of ALPs to nuclei and electrons, for example, translates given experimental macroscopic bounds to bounds on the coupling constants $g_S$, $g_P$, $g_V$, and $g_A$, which are stronger by a factor $1 + \rho_e/\rho_N$ than in the case of ALPs with an axion-like coupling (if we assume electric neutrality of the experimental source material).

Traditionally, scalar Yukawa-type interactions have been rather interpreted as corrections to Newtonian gravity, 
\begin{align}
    \label{eq:potential_Yukawa}
    V(r)=-\frac{G m_1 m_2}{r}\left(1+\alpha\,\re^{-r/\lambda}\right)=V_N(r)+V_Y(r)\>,
\end{align}
with couplings proportional to the masses $m_1$ and $m_2$, respectively. For such a mass-like coupling, we may neglect the interactions due to the electrons and obtain the induced total~pressure 
\begin{align}
\label{eq:yukawa_gs_relation}
    \mathcal{P}_Y^\text{total}(a) = -2\pi G m_N^2\alpha\lambda^2\Phi(\lambda)\>,
\end{align}
where $m_N$ is the nucleon mass, and $\rho_{U,i}$ and $\rho_{L,j}$ within the function $\Phi(\lambda)$ denote the number of nucleons per unit volume. This can directly be compared to an axion-like coupling to nucleons only, in which case $\rho_{U,i}$ and $\rho_{L,j}$ again denote the number of nucleons per unit volume. From $\mathcal{P}_Y^\text{total}(a) = \mathcal P_{SS,\text{axion}}^\text{total}(a)$, we obtain relation
\begin{align}
    \alpha = \frac{\hbar c}{4\pi G m_N^2}\,g_S^2\>.
\end{align}

If we consider an ALP-like coupling equal to nucleons and electrons, instead, the total pressure gets enhanced $\mathcal P_{SS,\text{ALP}}^\text{total}(a) = (1 + \rho_e/\rho_N)^2\,\mathcal P_{SS,\text{axion}}^\text{total}(a) = \mathcal{P}_Y^\text{total}(a)$, which provides the relation
\begin{align}
    \alpha = \frac{(1 + \rho_e/\rho_N)^2\hbar c}{4\pi G m_N^2}\,g_S^2\>,
\end{align}
assuming a constant ratio $\rho_e/\rho_N$ for each layer. For a mass-like coupling, \eqnref{eq:TP1} represents a generalization of the recently used expression~\cite{Klimchitskaya:2019fsm} that first appeared in this form in Ref.~\cite{Bordag:1999gx}.

In \figref{fig:yukawa}, we give an updated prospective exclusion graph, taking into account the error estimation discussed in \secref{sec:des:total}.
\begin{figure}[!ht]
    \includegraphics[scale=0.86]{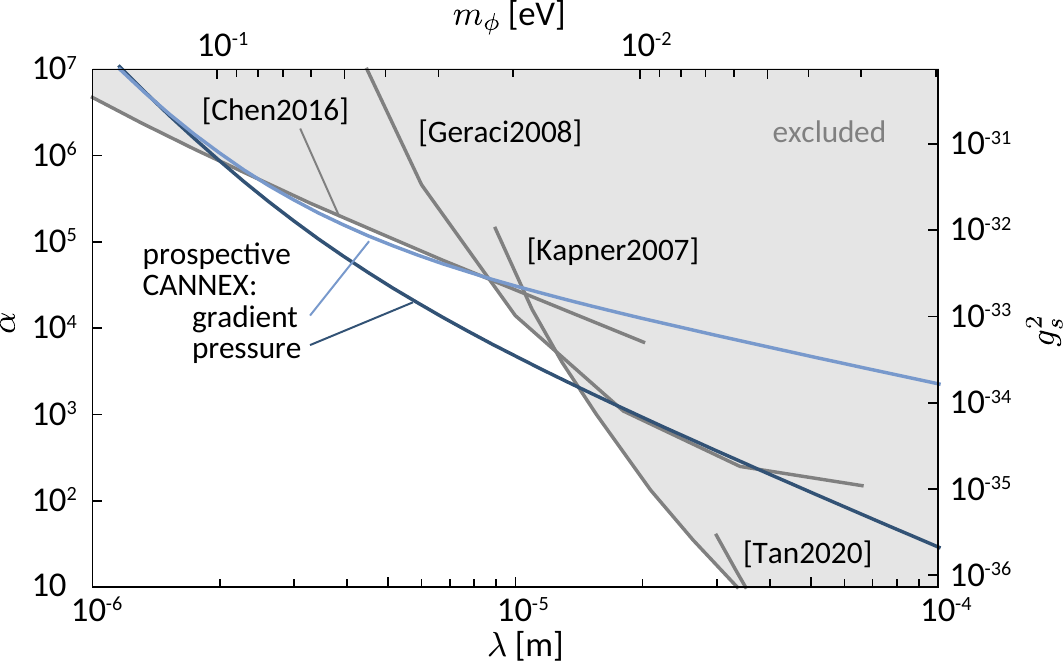}
    \caption{Prospective limits at 68\% confidence level (C.L.) on scalar Yukawa interactions in comparison with existing limits at 95\% C.L. from a recent Casimir experiment~\cite{Chen:2014oda} [Chen2016], Cavendish-type micro-oscillator~\cite{Geraci:2008hb} [Geraci2008], and torsion balances~\cite{Kapner:2006si,Tan:2020vpf} [Kapner2007] and [Tan2020], respectively. Scales for the generic coupling constant $g_S^2$ and the traditional mass-coupling constant $\alpha$ are related according to \eqnref{eq:yukawa_gs_relation} and indicated on the right and left axes, respectively.\label{fig:yukawa}}
\end{figure}
From \eqnref{eq:gasp}, one can obtain rough estimates concerning the relative importance of the specific interaction strengths in the case of axions and ALPs. Typically, the plates of the experimental setup consist of unpolarized matter. In this case, the effective potential \eqnref{eq:pseudoscalar-scalar} disappears after averaging over the macroscopic plates, while \eqnref{eq:scalar-scalar} gives the main contribution~\cite{Decca:2003td,Decca:2005yk,Decca:2007jq,Masuda:2009vu,Sushkov:2011md,Klimchitskaya:2012pc,Chen:2014oda}. An effective potential including the pseudo-scalar coupling $g_P$ and providing a non-vanishing contribution is to leading order given by a two scalar particle exchange~\cite{Adelberger:2003ph} and, hence, proportional to $g_P^4$~\cite{Bezerra:2014dja,Klimchitskaya:2019fsm} or $g_S^2g_P^2$. Indeed, using \eqnref{eq:gasp} and $m_q \sim 1$ MeV, we find for an axion that the `$g_P^4$' effective potential is proportional to $(g_P^q)^4 \propto \mathcal O(m_q^4/f_a^4) \sim 10^{-48}$--$10^{-60}$, while the effective `$g_P^2g_S^2$' potential is proportional to $(g_P^q)^2(g_S^q)^2 \propto \mathcal O(\theta_{\text{eff}}^2m_q^4/f_a^4) \sim 10^{-70}$--$10^{-82}$, where we have employed the experimentally determined upper value $\theta_{\text{eff}} \sim 10^{-11}$. Hence, we find that the latter potential is suppressed by a factor $(g_S^q/g_P^q)^2 = \theta_{\text{eff}}^2$ relative to the `$g_P^4$' potential
and, as such, is expectedly of far less relevance. This is in stark contrast to the situation for ALPs, for which $g_S$ and $g_P$ are a priori unrelated parameters. Indeed, $g_S$ is not expected to be suppressed by many orders of magnitude with respect to $g_P$. In this case, the couplings $g_P^4$ and $g_P^2g_S^2$ may be of similar magnitude, while the corresponding potentials should be of comparable importance.

In addition, in Reference \cite{Pitschmann:2020ejb}, a proposal for an extension of \cannex{} is investigated in which the lower plate is coated with a \SI{1}{\micro\metre} thick layer of ferromagnetic nickel. In this case, polarization can be achieved in situ by means of a magnetic field. After the field has been driven up to \SI{50}{\milli\tesla} to reach saturation of the nickel layer, it is subsequently reduced to zero. The nuclear and electron spins keep their remnant 
degrees of polarization, thereby allowing to measure spin-dependent one-scalar exchange interactions, e.g., proportional to $g_Sg_P$, as described by the effective potential \eqnref{eq:pseudoscalar-scalar}. For axions, the latter is proportional to $g_S^qg_P^q \propto \mathcal O(\theta_{\text{eff}}\,m_q^2/f_a^2) \lesssim 10^{-35}$--$10^{-41}$. Comparing this to the effective potential proportional to $(g_S^q)^2 \propto \mathcal O(\theta_{\text{eff}}^2\,m_q^2/f_a^2) \sim 10^{-46}$ -- $10^{-52}$, we find that the latter potential is suppressed by a factor $g_S^q/g_P^q = \theta_{\text{eff}}$ relative to the `$g_Sg_P$' potential. Again, for ALPs, the situation is different, and one may expect to be able to provide competitive bounds for the two potentials \eqnsref{eq:scalar-scalar} and (\ref{eq:pseudoscalar-scalar}).

Note that, as described in Section \ref{sec:des:det}, in order to obtain the limits in \figref{fig:yukawa}, we have to subtract the pressure (or pressure gradient) of the electrostatic, as well as the Casimir interaction. In this respect, a relative measurement, such as described in Refs.~\cite{Decca:2005qz,Chen:2014oda,Bimonte:2016}, is advantageous, as the uncertainty in the theoretical description of the dielectric functions is eliminated. However, given that \cannex{} aims to cover a wide range of distances, and considering that the proposed hypothetical interactions have a quite different dependence on distance than the Casimir interactions, a clear distinction still seems possible. Note also that, for hypothetical DE fields mentioned below, this problem does not exist, as a relative measurement can be realized in \cannex{} by changing the ambient matter density, as suggested earlier~\cite{Brax:2010xx,Almasi:2015zpa}.

Concerning the influence of curvature of the plates on the evaluation of prospective limits on Yukawa-like interactions, we refer to Section \ref{sec:des:flat}.
%
\subsection{Dark Energy}
\label{subsec:darkenergy}
As described in \secref{sec:ODSIDM}, a recent `screened' dark energy model is the symmetron model, for which the effective potential is such that it acquires a nonzero vacuum expectation value (VEV) in low-density regions, where the field spontaneously breaks symmetry, couples to matter and mediates a fifth force. However, in high density regions, the symmetry is restored, and the field disappears, which renders the interaction invisible to any observation or measurement. For \cannex{}, knowledge of the respective field profiles depending on the density and geometry of the plates is crucial. For the `$1$-dimensional approximation' of two plates of averaged density, finite thickness but infinite transversal extension exact solutions for symmetrons have been obtained in References~\cite{Brax:2017hna,Pitschmann:2020ejb} (for earlier approximate solutions, see references therein).
Similarly, for chameleons, the corresponding solutions have been obtained in Reference~\cite{Ivanov:2016rfs}.

\begin{figure}[!b]
    \includegraphics[width=120mm]{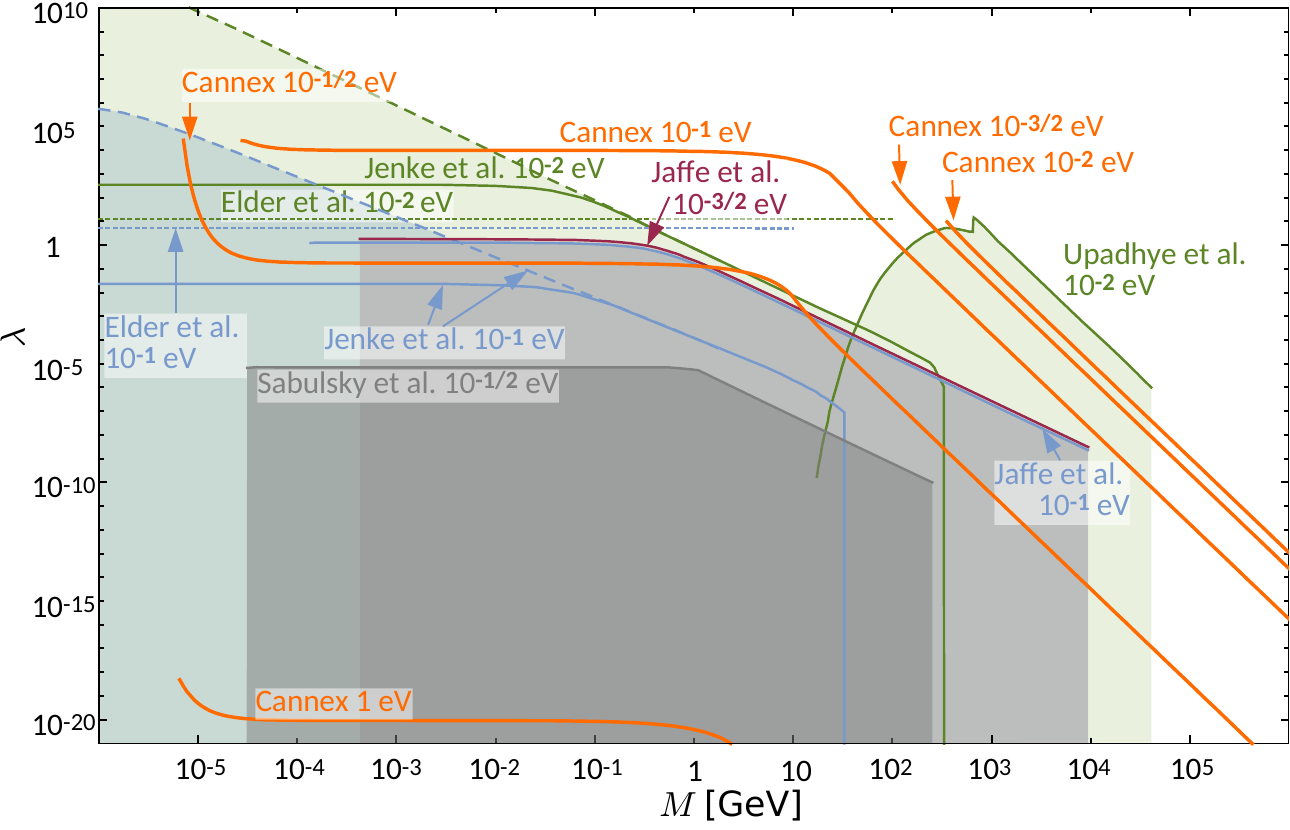}
    \caption{Limits on the symmetron parameter $\lambda$ as a function of $M$ at several fixed values of $\mu$. Existing limits from atom interferometry (Jaffe et al.~\cite{Jaffe:2016fsh}, 95\% C.L.; Sabulsky et al.~\cite{Sabulsky:2018jma}, 90\% C.L.), (prospective) Casimir experiments (Elder et al.~\cite{Elder:2019yyp}, 95\% C.L.), and torsion balance experiments (Upadhye et al.~\cite{Upadhye:2012rc}, no C.L. given). Limits for equal $\mu$ are shaded in blue: $10^{-2}\,$eV, red: $10^{-3/2}\,$eV, green: $10^{-1}\,$eV, and gray: $10^{-1/2}$\,eV. Prospective limits for \cannex{} (orange, 90\% C.L.) are computed for a separation of $a=\SI{10}{\micro\metre}$, based on the analysis of systematic effects in \secref{sec:design}, and include solutions for `broken and unbroken phase parameter ranges'~\cite{Brax:2017hna, Pitschmann:2020ejb}.\label{fig:symmetron_limits}}
\end{figure}

Whether the symmetron is in its symmetric phase inside the plates of \cannex{}, or in the symmetry broken phase depends not only on the environmental mass density but also on the specific parameter values of the symmetron. One has to distinguish two complementary cases. For parameter values with $\rho\geq \mu^2 M^2$, where $\rho$ denotes the averaged density of the \cannex{} plates, the symmetron is in its symmetric phase inside the plates but in the symmetry broken phase in the vacuum chamber of the experimental setup. However, for the complementary parameter values, the symmetron remains in its spontaneously broken phase (yielding a non-vanishing VEV), throughout the setup. Since the magnitude of the VEV is still smaller inside matter than in vacuum, a pressure is also induced on the plates in this case. This scenario has been investigated theoretically for the first time in Reference~\cite{Pitschmann:2020ejb}, where also the symmetron-induced pressure in the \cannex{} setup was investigated.
For parameter values for which the symmetron is in its symmetric phase inside the \cannex{} plates, the induced symmetron pressure was found to be
\begin{align}
\label{eq:symmetron_pressure}
  \mathcal P_\text{Sym} = -\frac{\rho_M}{4M^2}\frac{\mu_V^2}{\lambda}\frac{(k^2 - 1)^2}{\displaystyle1 + \frac{\rho_\text{eff}}{M^2\mu_V^2}}\>,
\end{align}
where $\rho_V\simeq0$ is the residual density of the vacuum surrounding the plates, $\rho_\text{eff} = \sqrt{\rho_M - M^2\mu^2}$, $\mu_V = \sqrt{\mu^2 - \rho_V/M^2} \simeq \mu$, and $k\in[0,1]$ denotes a parameter specific to the particular symmetron solution (see Reference~\cite{Brax:2017hna} for further details and definitions). For the complementary case with the symmetron in its spontaneously broken phase throughout, the induced pressure is given instead by~\cite{Pitschmann:2020ejb, Pitschmann:2020ejb}
\begin{align}
\label{eq:symmetron_pressureII}
  \mathcal P_\text{Sym} = \frac{\rho_M}{4\lambda M^2}\Big(\mu_M^2\big(2k_M^2 - 1\big) - \mu_V^2\Big)\>,
\end{align}
where $\mu_M = \sqrt{\mu^2 - \rho_M/M^2}$, and $k_M\in[0,1]$ is another parameter specific to the particular symmetron solution (see Reference~\cite{Pitschmann:2020ejb} for further details and definitions). \eqnsref{eq:symmetron_pressure} and \eqref{eq:symmetron_pressureII}, together, provide the pressure for the entire symmetron parameter space. 

It must be noted that, as found in Reference~\cite{Brax:2017hna,Pitschmann:2020ejb}, the symmetron admits not just one but rather a discrete spectrum of possible field solutions between two parallel plates. This multiplicity of solutions is reflected in discrete sets of possible values for the parameters $k$ and $k_M$ in \eqnsref{eq:symmetron_pressure} and \eqref{eq:symmetron_pressureII}, respectively. In the computation of prospective bounds on symmetron parameters from \cannex{}, we employ the particular solution without any nodes between the plates (fundamental mode). 

In Reference~\cite{Pitschmann:2020ejb}, the \cannex{} limits in Figure~\ref{fig:symmetron_limits} on the symmetron parameter $\lambda$ as a function of $M$ and for several discrete values of the parameter $\mu$ have been obtained and compared to results from other measurements. For these bounds for $\lambda(M)$ at 90\% confidence level, the predicted experimental sensitivity~\cite{Klimchitskaya:2019fsm} of \cannex{} (scaled by a factor 2) has been compared to numerical results obtained with \eqnsref{eq:symmetron_pressure} and \eqref{eq:symmetron_pressureII}.
Other laboratory constraints on symmetrons stem from atomic interferometry~\cite{Brax:2016wjk,Jaffe:2016fsh,Sabulsky:2018jma}, neutron gravity resonance spectroscopy~\cite{Cronenberg:2018qxf,Jenke:2020obe}, (prospective) Casimir experiments~\cite{Elder:2019yyp}, and torsion balances~\cite{Upadhye:2012qu}. One should also note that, for the \cannex{} setup, field solutions for the parameter values $\mu = 10^{-3/2}\,$eV and $\mu = 10^{-2}\,$eV exist only for the symmetron in its broken phase throughout. Therefore, the limits for these two parameter values are constrained by lower bounds in $M$ (see Ref.~\cite{Brax:2017hna}).

Concerning the influence of curvature on the plates on the evaluation of prospective limits on symmetrons, we refer to Section \ref{sec:des:flat}.
%
\section{Conclusions}
\label{sec:conclusion}
The Casimir And Non-Newtonian force EXperiment is one of very few instruments employing the geometry of macroscopic parallel plates to perform metrological force measurements. The concept was demonstrated to work as expected in a recent proof of principle experiment. However, technical and administrative problems prevented us from reaching the targeted sensitivity level. Based on our experience with the setup and on recent propositions to perform additional measurements, we now present a preliminary design for an improved setup. For the previously most influential systematic errors due to vibrations, thermal drift, and patch effects, we employ proven countermeasures. The seismic attenuation system is extended from one to two stages, which (according to the model) should result in an attenuation of $-$71\,dB and $-$106\, dB in amplitude in horizontal and vertical direction, respectively, around the sensor resonance frequency of roughly \SI{10}{\hertz}. The thermal stabilization concept was redesigned for local control of both the outer and the core chamber. In addition, the sensor is now actively stabilized to well within $1\,$mK around room temperature. The second (fixed) plate's temperature can be changed by roughly $\pm10\,^\circ$C to allow for measurements of forces out of thermal equilibrium. Finally, we include a recently found method to reduce the value of electrostatic patch forces by means of in-situ UV irradiation and subsequent Ar ion cleaning of the interacting surfaces.

\textls[-15]{Together, these improvements shall allow us to practically reach the targeted sensitivites of $1\,$nN/m$^2$ for pressures and $1\,$mN/m$^3$ for pressure gradients at plate \mbox{separations $\geq \SI{10}{\micro\metre}$}.} We support this claim by a detailed and transparent error estimation for all parts of the system. The most influential systematic errors at small separations are residual electrostatic patches and uncertainties in the large-scale deformation of our plates. At larger separations, for pressure measurements, the limitation comes from laser line width and electronic noise in the interferometric detection of the sensor extension. For the pressure gradient, the resolution and phase noise of the frequency-detecting PLL are limiting factors. While, at frequencies $>\SI{1}{\hertz}$, the new seismic attenuation system renders vibrations negligible, the long integration times of $\sim$$600\,$s introduce a sensitivity to ultra-low frequency earth tides and variations in the gravitational field. For this reason, we will use models and reference data from a nearby gravimetric laboratory to correct our data.

With the predicted sensitivity, the Casimir interaction can be measured for plate separations of 3--\SI{30}{\micro\metre} with high accuracy. Assuming this sensitivity to be realized, the difference between the theoretically predicted Casimir pressures (and gradients) obtained using the conflicting plasma and Drude approaches amounts to more than 100 times the experimental error in a large range of separations around the thermal wavelength, which would enable us to to clarify the r\^ole of dissipation for the thermal Casimir effect. Similarly, with a difference of more than 30 times the experimental error, we could clearly distinguish pressures and pressure gradients predicted for a non-local extension of the Drude model and the plasma model, which both lie within the errors of presently available data. We could also perform a measurement of non-equilibrium thermal Casimir forces and radiative heat transfer with up to $10~^\circ$C difference between the two plates.

With respect to hypothetical dark sector interactions, \cannex{} may contribute to limits on scalar, pseudoscalar, and vector dark matter interactions, as well symmetron and chameleon dark energy interactions. The latter model could be excluded entirely as candidate for dark energy. Eventually, for pressures, measurements at the level of the gravitational attraction of the two plates seem possible.

After a long development and testing phase, thus, \cannex{} may finally serve as a metrology platform for interfacial and gravity-like interactions. It could provide experimental input for the fields of Casimir physics, as well as the physics of the `dark sector', and answer long-standing questions in these fields.

\authorcontributions{Investigation dark interactions: M.P.; Investigation experimental part, visualization: R.I.P.S.; all authors contributed equally to writing. All authors have read and agreed to the published version of the manuscript.}

\funding{This research received Open Access Funding by TU Wien.} 

\institutionalreview{Not applicable.}

\informedconsent{Not applicable.}

\dataavailability{The experimental data and results from calculations can be requested from the corresponding author.}

\acknowledgments{We thankfully acknowledge support in the mechanical design from Ivica Galic and Max Maichanitsch.}

\conflictsofinterest{The authors declare no conflict of interest.} 

\appendixtitles{yes} 
\appendixstart
\appendix

\section{Errors Due to Vibration}
\label{app:vibr_errors}
We start by modeling the vertical response of the sensor $z_s(t)$ to vibrations of its frame (core chamber) $z_c(t)$ and pressures $p_s(t)$. We use a lumped parameter model for a simple harmonic oscillator with viscous and internal dampings $\gamma=m_\text{eff}\omega_0/Q$ and $\phi$, respectively.
\begin{align}
    m_\text{eff}\ddot{z}_s(t)+\ri m_\text{eff}\omega_0/Q\left[\dot{z}_s(t)-\dot{z}_c(t)\right]+k(1+\ri\phi)\left[z_s(t)-z_c(t)\right]=Ap_s(t)\>.
\end{align}

Assuming purely sinusoidal excitation, this yields for the transfer functions in frequency~space 
\end{paracol}
\nointerlineskip
\appendix
\begin{align}
    T_\text{cs,rel}(\omega)=\frac{\tilde{z}_s(\omega)}{\tilde{z}_c(\omega)}=\frac{\omega^2}{\omega_0^2(1\!+\!\ri\phi)-\omega^2+\ri\omega\omega_0/Q}\>,\quad\text{and }T_{Ps,rel}=\frac{\tilde{z}_s(\omega)}{\tilde{P}_s(\omega)}=\frac{A T_\text{cs,rel}(\omega)}{m_\text{eff}\omega^2}\,.\label{eq:tf_sensor}
\end{align}
\begin{paracol}{2}
\switchcolumn

Here, the variables with tilde are the Fourier-transformed versions of the respective quantities. For low frequencies, the loss angle $\phi$ is in the $10^{-5}$ range~\cite{AlShourbagy:2006vf} and can safely be neglected for the following error estimation. In the precise calibration of $\omega_0$, however, $\phi$ must be taken into account.
For interactions depending on the separation, we have
\begin{align}
    M(a+\delta a)\approx M(a)+\delta a\,M'(a)+\inv{2}\delta a^2 M''(a)\>,\label{eq:vibr_error_expansion}
\end{align} 
with $M(a)$ being either the relevant pressure $P(a)$ or pressure gradient $P'(a)$. If we assume $\delta a$ to be purely stochastic with $\langle\delta a\rangle=0$, the second term on the right side vanishes, and the third one gives an offset $\delta M_{nl}=(1/2)\delta a^2 M''(a)$ that we refer to as the non-linearity error in $M(a)$. By demanding $\delta M_\text{nl}<M_\text{lim}$, we obtain the max. sensor vibration amplitude $\delta a_\text{max,nl}=\sqrt{2 |M_\text{lim}/M''(a)|}$. In order to translate the sensor movement to vibrations $\delta z_c$ of the core chamber, we use $\delta z_{c,\text{max,nl}}(\omega)=\delta a_\text{max,nl} |T_\text{cs,rel}|^{-1}(\omega)$. Note that the non-linearity depends on the total vibration amplitude at all frequencies for which the spectral limit $\delta z_{c,\text{max}}(\omega)$ depends on the frequency-dependent response of the sensor. The systematic error $\delta M_\text{nl,rms}$ is obtained from the (measured or, here, computed) core vibration amplitude $z_c(\omega)$ by integrating over the entire frequency space.
\begin{align}
\delta M_\text{nl,rms}=M''(a)\left[\int\limits_0^\infty\!{\rm d}\omega\,\left(\inv{2}\delta z_c^2(\omega) |T_\text{cs,rel}(\omega)| \right)^2\right]^{1/2}\,.
\end{align}

Apart from the non-linearity, the distance dependence of $M(a)$ also introduces a statistical RMS error via the linear (second) term in \eqnref{eq:vibr_error_expansion}:
\begin{align}
   \delta M_{a,\text{rms}}= M'(a)\left[\int\limits_\text{BW}\!{\rm d}\omega\, \left(\delta z_c(\omega) |T_\text{cs,rel}(\omega)| \right)^2\right]^{1/2}=\delta a_{rms}M'(a)\,.\label{eq:err_vibr_rms}
\end{align}

Here, BW indicates the relevant bandwidth. For pressure measurements integrating over a time $\tau_m$, we have to consider the frequency range $\omega\in[0,\,\tau_m^{-1}]$. For pressure gradient measurements, only the range $\omega\in[\omega_0- \omega_\text{BW}/2,\,\omega_0+ \omega_\text{BW}/2]$ around the mechanical resonance $\omega_0$ of the sensor is relevant, and $\omega_\text{BW}\approx 2\pi\times\SI{3}{\milli\hertz}$. In order to derive a limit for the permissible vibration $z_c(\omega)$, we set $\delta M_{a,\text{rms}}<M_\text{lim}$. Since both relevant bandwidths in \eqnref{eq:err_vibr_rms} are small, we approximate $\delta z_c(\omega)\approx\text{const.}$ within the integration range, yielding $\delta z_{c,\text{max,rms}}\approx M_\text{lim}/M'(a)[\int_\text{BW}\!{\rm d}\omega\,|T_\text{cs,rel}|^2]^{-1/2}$. Note that $\delta M_{a,\text{rms}}$ is purely statistical.

Yet another way to derive limits comes from the fact that extensions of the sensor are interpreted in terms of a pressure causing them, $\delta P_{P}=\delta z_c(\omega)|T_\text{cs,rel}(\omega)T_{Ps,rel}^{-1}(\omega)|$~\cite{Sedmik:2018kqt}. We immediately obtain $z_{c,\text{max,P}}(\omega)=P_\text{lim}|T_\text{cs,rel}(\omega)T_{Ps,rel}^{-1}(\omega)|^{-1}$. This limit does not depend on $a$ or the measured interaction but solely on the way pressures are measured. Integrating over $\omega\in [0,\tau_m^{-1}]$ yields another statistical error $\delta P_{P,\text{rms}}$.

Finally, we also consider phase noise for frequency shift measurements. We have the oscillation $a_{ex}(t)=|T_{Ps,rel}(\omega_0)|\vare_0 V_{ex}^2/(4 a^2)\sin(\omega_0 t+\varphi_{Ps})$ due to electrostatic excitation of the sensor with phase $\varphi_{Ps}=\arg T_{Ps,rel}(\omega_0)=-\pi/2$ being held constant by the PLL. We neglect for the moment any additional forces acting on the sensor. At the zero transition of the sine, we obtain the relation $\delta \varphi(t)\approx \left[\partial a_{ex}(t)/\partial \varphi_{Ps}\right]^{-1}\delta a$ for the maximum phase noise with amplitude $\delta\varphi$. Naturally, $\partial a_{ex}(t)/\partial \varphi_{Ps}=|T_{Ps,rel}|(\omega_0)\vare_0 V_{ex}^2/(4 a^2)=a_{ex}$ is the excitation amplitude. For this reason, we can write for the amplitudes $\delta\varphi=1/\text{SNR}$, where the SNR refers to the ratio between excitation signal $a_{ex}$ and vibration signal $\delta a$. We propagate the error further to frequency:
\begin{equation}
    \varphi_{Ps}(\omega)=\arctan{}\frac{\omega\omega_0}{Q(\omega^2-\omega_0^2)} \>,\quad\text{and } \delta\omega=\left[\diff{\varphi_{Ps}(\omega)}{\omega}\right]^{-1}\delta\varphi\,.
\end{equation}

Then, we interpret $\delta \omega$ as the error in the measured frequency, $\omega_m=\omega_0-\Delta \omega+\delta\omega$ given by \eqnref{eq:freq_shift}. Finally, from $\delta P'(a)=[\partial P'(a)/\partial \omega_m]\delta\omega$, we obtain 
\begin{align}
    \delta P'_{\varphi}(a)=\frac{m_\text{eff}\omega_0^2}{Q\,\text{SNR}}\>,
\end{align} as a further statistical error. We note that, in the same way, the (as well statistical) errors $\delta P'_\text{PLL}$ due to phase noise in the used lock-in amplifier, and $\delta P'_{Br}$ due to Brownian noise of the sensor, can be computed. The latter contributions, however, are negligible against the effect of vibrations. Limits on the permissible core vibration from phase noise can be obtained by fixing the desired SNR in $\delta z_{c,\text{max}}=P'_\text{lim} (a_{ex}/\text{SNR})\, Q/(m_\text{eff}\omega_0^2)\, T_\text{cs,rel}^{-1}$. Based on experience, we set for the calculation of limits $\text{SNR}=10$.
\begin{figure}[H]
    \mbox{}\includegraphics[scale=0.86]{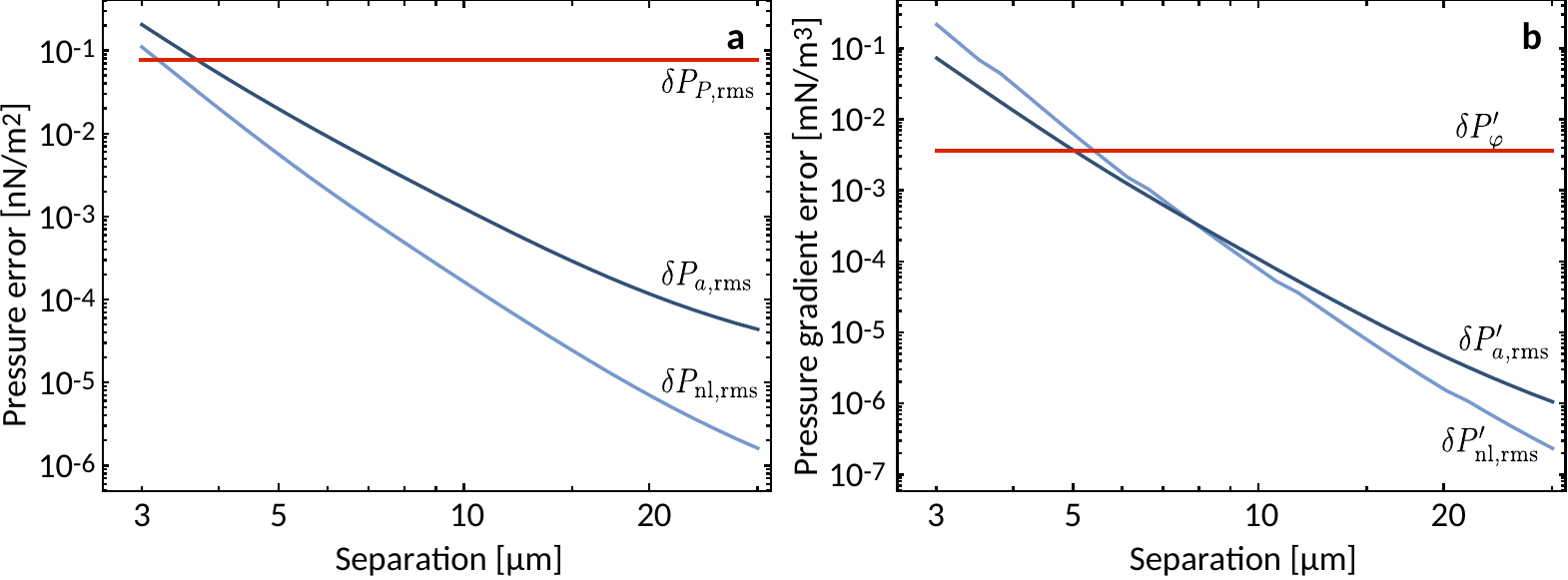}
    \caption{Amplitudes of the different contributions to seismic noise considering an integration time $\tau_m=600\,s$ and statistical averaging over 100 measurements for RMS and phase noise errors. (\textbf{a}): errors on the pressure. (\textbf{b}): errors on the pressure gradient.}
    \label{fig:seismic_contrib}
\end{figure}
\vspace{-6pt}
\begin{figure}[H]
    \includegraphics[scale=0.86]{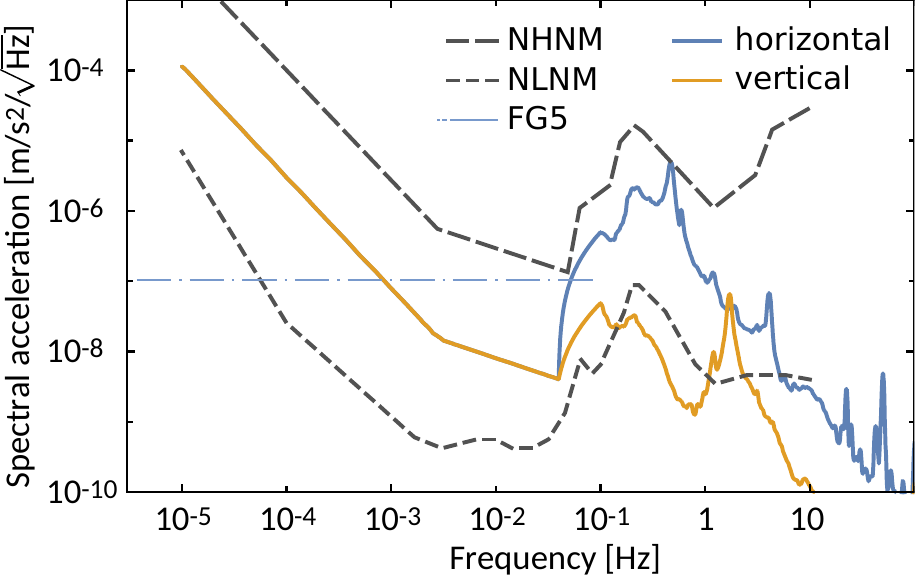}
    \caption{Extrapolation of spectral seismic acceleration below $0.1\,$Hz using the `new high noise model' (NHNM)~\cite{Peterson:1993}, scaled to fit the measured vertical acceleration at $0.1\,$Hz. We also indicate the noise level of FG-5 data recorded at a nearby station that can be used to correct our pressure data.\label{fig:seismic_low_freq}}
\end{figure}

In \figref{fig:seismic_contrib}, we show the amplitudes of the different contributions discussed above as a function of distance for $\tau_m=600\,$s and $\omega_\text{BW}=2\pi\, 3\,$mrad/s. All integrations are performed using the predicted attenuation levels shown in \figref{fig:seismic_performance} in the main text. Our geophone measurements, however, are dominated by electronic $1/f$ noise at frequencies $\lesssim 0.1\,$Hz. In order to perform the required integrations for $\delta M_{\rm nl}$, $\delta M_{\rm rms}$, and $\delta P'_\varphi$, therefore, we extrapolate our seismic data using a model for the global maximum vibration amplitude~\cite{Peterson:1993} at lower frequencies down to $10^{-5}\,$Hz (see \figref{fig:seismic_low_freq}). To be precise, we take the measured level of vertical seismic background at frequency $\SI{0.1}{\hertz}$ and scale the NHNM curve~\cite{Peterson:1993} such that it connects with this value. As in global measurements, horizontal noise generally turns out smaller than the vertical one, we take equal levels for horizontal and vertical noise as worst case. In principle, the errors $\delta P_{a,\text{rms}}$ and $\delta P_{P,\text{rms}}$ are purely statistical, which should allow us to decrease them by taking longer measurements. An unavoidable problem, however, are the Earth tides~\cite{Farrell:1973}, especially the semidiurnal mode $M_2$ with an amplitude of 64\,cm and period of 12.41\,h~\cite{Agnew:2010} (corresponding to a pressure amplitude of roughly 3.8\,nN/m$^2$ on the sensor). Therefore, statistical reduction of $P_{a,\text{rms}}$ and $\delta P_{P,\text{rms}}$ affected by these large vibrations at very low frequency, would require measurements over several days. Instead, since the setup is stabilized thermally (see \secref{sec:des:thermal}), the actual influence of the Earth tides at the location of the experiment over the day cycle can be measured and compared to localized models of the tides, which, in turn, can be used to correct our pressure data. Finally, near the position of \cannex{}, a geophysical laboratory measures the tides using an FG-5 gravimeter. Using this data, data correction could be achieved at the $10^{-8}\,{\rm m/s}^2$-level. Any residual stochastic very-low-frequency noise can be reduced by averaging measurements. We, therefore, cut off all integrations regarding pressures at $10^{-4}\,$Hz and consider a reduction factor $1/\sqrt{N}$ for $N=100$ measurements in the given limits on $P_{a,\text{rms}}$ and $\delta P_{P,\text{rms}}$. The same averaging is applied in the computation of the error $\delta P'_{a,\text{rms}}$. 
Note that this very-low-frequency contribution was not considered in the error budget in Ref.~\cite{Klimchitskaya:2019fsm}. We, consequently, add the different contributions to arrive at total statistical and systematic errors
$\delta P_\text{seis,stat}=[\delta P_{a\text{,rms}}^2+\delta P_{P,\text{rms}}^2]^{1/2}$, $\delta P'_\text{seis,stat}=[\delta {P'}_{a\text{,rms}}^2 +{P'}_{\varphi}^2 ]^{1/2}$, and $\delta M_\text{seis,sys}=M_\text{nl,rms}$.
%
%
\end{paracol}
\reftitle{References}

\end{document}